\documentclass[preprint, flushrt]{aastex}

\newcommand{\etal}{\textrm{et al. }}

\newcommand{\eg}{\textrm{e.g. }}

\newcommand{\Ha}        {\,{\rm {H\alpha}}}
\newcommand{\Hb}        {\,{\rm {H\beta}}}

\catcode`\@=11 % This allows us to modify PLAIN macros.
\def\lsim{\mathrel{\mathpalette\@versim<}}
\def\gsim{\mathrel{\mathpalette\@versim>}}
\def\@versim#1#2{\vcenter{\offinterlineskip
        \ialign{$\m@th#1\hfil##\hfil$\crcr#2\crcr\sim\crcr } }}

\begin{document}

\title{Active Galactic Nuclei in the Sloan Digital Sky Survey: II. Emission-Line Luminosity Function}
\author{
Lei~Hao\altaffilmark{1,2}, 
Michael~A.~Strauss\altaffilmark{1}, 
Xiaohui~Fan\altaffilmark{3}, 
Christy~A.~Tremonti\altaffilmark{3}, 
David~J.~Schlegel\altaffilmark{1}, 
Timothy~M.~Heckman\altaffilmark{4}, 
Guinevere~Kauffmann\altaffilmark{5}, 
Michael~R.~Blanton\altaffilmark{6}, 
James~E.~Gunn\altaffilmark{1},
Patrick~B.~Hall\altaffilmark{1}, 
\v{Z}eljko~Ivezi\'{c}\altaffilmark{1}, 
Gillian~R.~Knapp\altaffilmark{1}, 
Julian~H.~Krolik\altaffilmark{4},
Robert~H.~Lupton\altaffilmark{1}, 
Gordon~T.~Richards\altaffilmark{1},
Donald~P.~Schneider\altaffilmark{7},
Iskra~V.~Strateva\altaffilmark{1}, 
Nadia~L.~Zakamska\altaffilmark{1},
J.~Brinkmann\altaffilmark{8},
Gyula~P.~Szokoly\altaffilmark{5}}
\altaffiltext{1}{Princeton University Observatory, Princeton, NJ 08544}
\altaffiltext{2}{Current address: Astronomy Department, Cornell University, Ithaca, NY 14853; haol@isc.astro.cornell.edu}
\altaffiltext{3}{Steward Observatory, University of Arizona, 933 North Cherry Avenue, Tucson, AZ 85721}
\altaffiltext{4}{Department of Physics and Astronomy, Johns Hopkins University, 3400 North Charles Street, Baltimore, MD 21218} 
\altaffiltext{5}{Max-Planck Institut f\"{u}r Astrophysik, D-85748 Garching, Germany}
\altaffiltext{6}{Center for Cosmology and Particle Physics, Department
of Physics, New York University, 4 Washington Place, New York, NY
10003}
\altaffiltext{7}{Department of Astronomy and Astrophysics, Pennsylvania State University, University Park, PA 16802}
\altaffiltext{8}{Apache Point Observatory, P.O. Box 59, Sunspot, NM 88349-0059.}

\slugcomment{\textit{\today}}

\begin{abstract}
The emission line luminosity function of active galactic nuclei (AGN)
is measured from about 3000 AGN included in the main galaxy sample of
the Sloan Digital Sky Survey within a redshift range of
$0<z<0.15$. The $\Ha$ and [OIII]$\lambda 5007$ luminosity functions
for Seyferts cover luminosity range of $10^{5-9}$$L_\odot$ in
H$\alpha$ and the shapes are well fit by broken power laws, without a
turnover at fainter nuclear luminosities. Assuming a universal
conversion from emission line strength to continuum luminosity, the
inferred B band magnitude luminosity function is comparable both to
the AGN luminosity function of previous studies and to the low
redshift quasar luminosity function derived from the 2dF redshift
survey. The inferred AGN number density is approximately 1/5 of all
galaxies and about $6\times 10^{-3}$ of the total light of galaxies in the $r$-band comes from the nuclear activity. The numbers of Seyfert 1s and Seyfert
2s are comparable at low luminosity, while at high luminosity, Seyfert
1s outnumber Seyfert 2s by a factor of 2-4. In making the luminosity
function measurements, we assumed that the nuclear luminosity is
independent of the host galaxy luminosity, an assumption we test {\it
a posteriori}, and show to be consistent with the data. Given the
relationship between black hole mass and host galaxy bulge luminosity,
the lack of correlation between nuclear and host luminosity suggests
that the main variable that determines the AGN luminosity is the
Eddington ratio, not the black hole mass. This appears to be different
from luminous quasars, which are most likely to be shining near the
Eddington limit.

\end{abstract}


\section{Introduction}
Active Galactic Nuclei (AGN), including high-luminosity quasars and
low-luminosity Seyferts or LINERs, are among the most intriguing
objects in the universe. The optical luminosity functions of AGN
overall, as well as different types of AGN, hold important clues about
the demographics of the AGN population, which in turn provide strong
constraints on physical models and evolutionary theories of AGN. Many
attempts have been made to derive the luminosity function of quasars
(Schmidt \& Green 1983; Marshall \etal 1983; Boyle \etal 1988, 1990,
2000; Hewett, Foltz \& Chaffee 1993; Schmidt, Schneider \& Gunn 1995;
Hawkins \& Veron 1995; Koehler \etal 1997; Goldschmidt \& Miller 1998;
La Franca \& Cristiani 1997, 1998; Fan \etal 2001). The largest study
to date is that of Croom \etal (2004), who evaluate the QSO luminosity
function and its cosmological evolution from the combined 2dF QSO
Redshift Survey and 6dF QSO Redshift Survey. Their luminosity function
covers a luminosity range of $M_B<-22.5$(with $H_0=100 {\rm km s^{-1}
\ Mpc^{-1}}$) and redshift range of $0.4<z<2.1$. However, selecting
AGN via their colors requires that the nuclear luminosity be at least
comparable to the host galaxy luminosity to be detected. Therefore
this method is biased against low luminosity AGN.

To extend the AGN optical luminosity function to low luminosities,
where the contribution of the host galaxy can equal or exceed that of
the AGN, we must select AGN via their spectroscopic features in a
galaxy redshift survey. Huchra \& Burg (1992) selected 25 Seyfert 1s
and 23 Seyfert 2s from the CfA redshift survey (Huchra \etal 1983),
and used these AGN to measure their luminosity functions. More
recently, Ho \etal (1997a) carried out a careful and uniform
spectroscopic survey of 486 nearby galaxies selected from the RSA
catalog (Sandage \& Tammann 1981) using very small apertures centered
on the nuclei. From these galaxies they found 211 AGN, including 94
LINERs, 65 transition objects and 52 Seyferts. Ulvestad \& Ho (2001)
obtained an optical AGN luminosity function from their
sample. Vila-Vilaro (2000) used this sample as well as the starforming
galaxy sample to estimate the shape of the luminosity functions for
Seyferts, LINERs, transition objects and starforming galaxies. The
luminosity functions obtained from these analyses extended to much
lower AGN luminosities, $M_B \sim -17$, than that obtained from
color-selected AGN samples, but they suffer from two disadvantages: 1)
the AGN samples are small, therefore the luminosity functions obtained
from these samples are subject to large uncertainties, and 2) previous
studies all used the B band magnitude of the entire galaxy instead of
the nuclear luminosity as the variable in their luminosity
functions. As a result, the luminosity functions are severely
contaminated by the AGN host galaxies. In order to overcome these
disadvantages, large redshift surveys and careful evaluation of the
luminosity functions are needed. The Sloan Digital Sky Survey (SDSS),
with its large number of high quality spectra available, provides us a
great opportunity to finally fulfill this.

The SDSS (York \etal 2000; Stoughton \etal 2002; Abazajian \etal 2003,
2004) is an imaging (Gunn \etal 1998) and spectroscopic survey that
will eventually cover approximately one-quarter of the Celestial
Sphere and collect spectra of $\sim 10^6$ galaxies and $10^5$
quasars. Software pipelines automatically perform the data processing:
astrometry (Pier \etal 2003); source identification, deblending and
photometry (Lupton \etal 2001); photometric calibration (Fukugita
\etal 1996; Smith \etal 2002); spectroscopic target selection
(Eisenstein \etal 2001; Strauss \etal 2002; Richards \etal 2002);
spectroscopic fiber placement (Blanton \etal 2003a) and spectroscopic
data reduction. The SDSS spectra are of high quality with spectral
resolution of about 2000 and typical signal to noise ratio of 16 per
pixel for the galaxy sample. This allows construction of a uniform
large sample of AGN identified from their spectroscopic features.

Hao \etal (2005) (Paper I) described the selection of AGN from the
SDSS spectroscopic data. We start from a low-redshift ($z<0.33$)
galaxy sample that is complete in $r$ band Petrosian magnitude at
17.77 (Strauss \etal 2002) over 1151 square degrees (see Zakamska
\etal 2003 for a selection of high redshift type II quasars from the
SDSS). To properly measure the emission-line properties of these
galaxies, we applied a stellar-subtraction procedure to each galaxy in
the sample with a set of absorption templates developed from a
Principal Component Analysis (PCA) of SDSS absorption-line
galaxies. The broad-line AGN are identified by their broad $\Ha$
emission line width: they all satisfy Full Width at Half Maxima of
$\Ha$ (FWHM($\Ha$))$>$1,200 km s$^{-1}$, a natural break point in the
distribution of $\Ha$ line widths. The narrow-line AGN are selected
via their locations in the emission-line ratio diagnostic diagrams
(Veilleux \& Osterbrock 1987). Kewley \etal (2001) developed
theoretical upper-limits for starburst galaxies in the diagrams and
proposed them to be used as the criteria to separate narrow-line AGN
from the usual star-forming galaxies. Kauffmann \etal (2003) plot the
SDSS galaxies in the [OIII]/$\Hb$ $\sim$ [NII]/$\Ha$ diagram and
noticed that the galaxies naturally separate into two branches. They
suggest classifying the galaxies in the second branch with higher
[NII]/$\Ha$ value as AGN. The separation point of the two branches
lies well below Kewley's separation line in the diagram. Therefore,
the Kauffmann criterion selects many more narrow-line AGN than
Kewley's criteria. In Paper I, we applied both criteria and compiled
an AGN sample containing 1317 broad-line AGN, 3074 narrow-line AGN via
Kewley's separation and 10,700 narrow-line AGN via Kauffmann's
criteria. Because they are basically low-luminosity AGN, in this paper
we will sometimes call these AGN Seyferts and refer to the broad-line
AGN sample as the Seyfert 1 sample, the narrow-line AGN sample
selected via Kewley's separation as the Seyfert 2 sample and that
selected via Kauffmann's criteria as the Seyfert 2$^*$ sample.

To measure the luminosity function, we first need to evaluate the
selection function; this is discussed in $\S2$. We define our emission
line AGN luminosity functions and describe our method to measure them
in $\S3$. In $\S4$ we show the luminosity function results. In $\S5$,
we apply several {\it a posteriori} checks on our luminosity function
results and the assumptions used in deriving them. We further compare
our results with the AGN luminosity functions obtained from previous
studies in $\S6$. In $\S7$ we discuss issues related to the reddening
correction and we summarize in $\S8$. Throughout this paper, we will
be using a cosmology with $(\Omega_m, \Omega_\Lambda)=(0.3,0.7)$ and
$H_0=100 {\rm km s^{-1} \ Mpc^{-1}}$.

\section{Selection Function}
SDSS spectra are taken with a fixed $3''$ aperture, thus the observed
spectra represent the sum of both the nuclear light and part or all of
the host galaxy light. The continuum component of the low-luminosity
AGN are difficult to separate from the host galaxy spectrum. Therefore
we choose to use the emission-line luminosities to represent the
nuclear luminosities and as variables in the AGN luminosity
function. Emission lines, such as [OIII], $\Ha$ and [OII] are good
representatives of nuclear activity, except that in some galaxies the
emission lines also include a contribution from star-forming
activity. As explained in Paper I and further in Section 4, this
contamination is not severe for AGN identified via Kewley's criteria,
since they are galaxies that are dominated by AGN components. AGN
selected via Kauffmann's criteria, however, might suffer this
contamination since many of them are AGN + starburst galaxies. As
demonstrated in Kauffmann \etal (2003), the [OIII] emission lines for
these galaxies, however, include little contamination from star
formation. Thus, the [OIII] luminosity function derived from the
narrow-line AGN sample using the Kauffmann \etal criteria will also be
a good indicator of AGN activity. In this paper, we will obtain the
$\Ha$, [OIII] and [OII] luminosity functions for samples obtained by
the criteria both of Kewley \etal (2001) and Kauffmann \etal (2003).

As described in Paper I and Section 1, the selection criteria for our
AGN include both an overall galaxy magnitude cut and selection via
certain spectral features. Thus the detection probability will depend
not only on the emission line luminosity, but also the spectral
features of the host galaxy, associated noise, and the overall galaxy
luminosity, which are different for each AGN. Therefore we will use
Monte-Carlo simulations that include all these factors to obtain the
detection probability function for each AGN.

The basic idea of the simulation is to evaluate the probability of
classifying a given object in our sample as an AGN, as a function of
the nuclear luminosity. We make a simplifying assumption to allow us
to continue: we assume that the nuclear luminosity of an AGN (as
represented by emission-line luminosity) is independent of its host
galaxy luminosity. Therefore, as the nuclear luminosity changes in the
simulation, the properties of the host galaxy and its associated noise
are kept unchanged. This assumption is not obviously true and has
significant physical implications of its own. We will show in $\S5$
that our data are consistent with this assumption.

Based on the selection criteria described above, there are several
possible reasons that an AGN will not be identified as the nuclear
luminosity is decreased. First, if the nuclear luminosity is too weak,
the emission lines will be undetectable. Second, as the emission line
strength is decreased, the noise in the measurement of the emission
line ratios and the $\Ha$ emission line width will increase. For AGN
near the limit of the selection criteria, this noise will sometimes
cause them to be no longer identified as AGN. Third, as the nuclear
luminosity is decreased, the overall flux of the galaxy, which is the
sum of contributions from the nucleus and the host galaxy, is
decreased as well. When the total brightness reaches the galaxy target
magnitude limit ( $r$ band Petrosian magnitude equal to 17.77) the AGN
will drop out of our sample. Therefore, the detection probability
increases as the nuclear luminosity increases. The detailed behavior
depends on the spectral features of individual spectra, which are
different from object to object. Therefore our simulations are carried
out on an object to object basis.

If we denote the original spectrum of an AGN observed in a $3''$
aperture as $f(\lambda)$ and the emission line spectrum obtained after
the continuum subtraction (Paper I) as $f_e'(\lambda)$, then
we can build a pure emission-line spectrum $f_e(\lambda)$ from
$f_e'(\lambda)$ that contains only the emission lines relevant to the
AGN identification by multiplying $f_e'(\lambda)$ by a window function: 
\begin{equation}
  W(\lambda)= \left\{ \begin{array}{ll}
               1  & \mbox{if $|\lambda-\lambda_k|<A_k$ for any $k$} \\
               0  & \mbox{otherwise}  
               \end{array}
       \right .     
\label{eq:window}
\end{equation}
Here, the $\lambda_k$ are the central wavelengths corresponding to
$\Ha$, H$\beta$, [OIII]$\lambda 4959$ [OIII]$\lambda 5008$,
[NII]$\lambda\lambda6548,84$, [SII]$\lambda\lambda6719,32$ and
[OI]$\lambda 6300$. $A_k = 3\sigma_k$, where $\sigma_k$ is the
Gaussian width of the line $k$. If two components are preferred in the
fitting for $\Ha$ or $\Hb$, we use the width of the broader component
(Paper I).

We use $f_e(\lambda)$ to represent the nuclear spectrum and
$f_a(\lambda)=f(\lambda)-f_e(\lambda)$ represents the host galaxy
spectrum. During the simulation, we evaluate the corresponding
detection probability while changing the nuclear luminosity, i.e., the
flux of the emission line spectrum $f_e(\lambda)$, by scaling by a
constant $c<1$. Based on our assumption of no correlation between AGN
and galaxy luminosity, the host galaxy spectrum $f_a(\lambda)$ and the
associated noise should be kept unchanged. However, when
$f_e(\lambda)$ is decreased, the noise within the relevant emission
line range is scaled down accordingly, even though the host galaxy,
which may dominate the spectrum, is unchanged. Therefore to keep the
noise within the emission line range unchanged, we generate a noise
spectrum. The noise associated with each wavelength pixel is randomly
generated following a Gaussian distribution. The amplitude of the
noise is chosen to be comparable to the noise of the original spectrum
at each pixel. The noise spectrum is multiplied by the emission-line
window function of Equation~\ref{eq:window}, thus the host galaxy
spectrum outside of the emission-line range is unchanged. If we denote
the final noise spectrum as $f_{noise}(\lambda)$, we can build a new
spectrum with decreased nuclear luminosity and emission line strength
as:
$$f_{new}(\lambda)=f_a(\lambda)+f_{noise}(\lambda)+c\cdot f_e(\lambda)$$

The whole process of AGN identification, including stellar
subtraction, is applied to this new spectrum to check if it still
identified as an AGN, i.e., for broad-line AGN, if its $\Ha$ emission
line has FWHM larger than 1,200 km s$^{-1}$ and for narrow-line AGN,
if its emission-line ratios satisfy the AGN emission line ratio
criteria (Kewley \etal 2001; Kauffmann \etal 2003; Paper I).

For each specific nuclear luminosity tested, i.e., for each value of
$c$, the whole process is repeated ten times, each with a different
randomly generated noise spectrum $f_{noise}(\lambda)$. The detection
probability associated with this nuclear luminosity and therefore this
coefficient $c$ is given by the number of times the new spectrum is
identified as an AGN, divided by the total number of tests, namely
ten.

The detection probability as a function of nuclear luminosity
therefore can be obtained by carrying out this simulation for a series
of nuclear luminosities, i.e., a series of coefficients $c$. The
values of $c$ are chosen to allow luminosities between $\sim 10^5
L_\odot$ and $\sim 10^9 L_\odot$ in logarithmic steps of $10^{0.034}$.
 
\begin{figure}[t]
\centerline{\includegraphics[angle=0, width=\hsize]{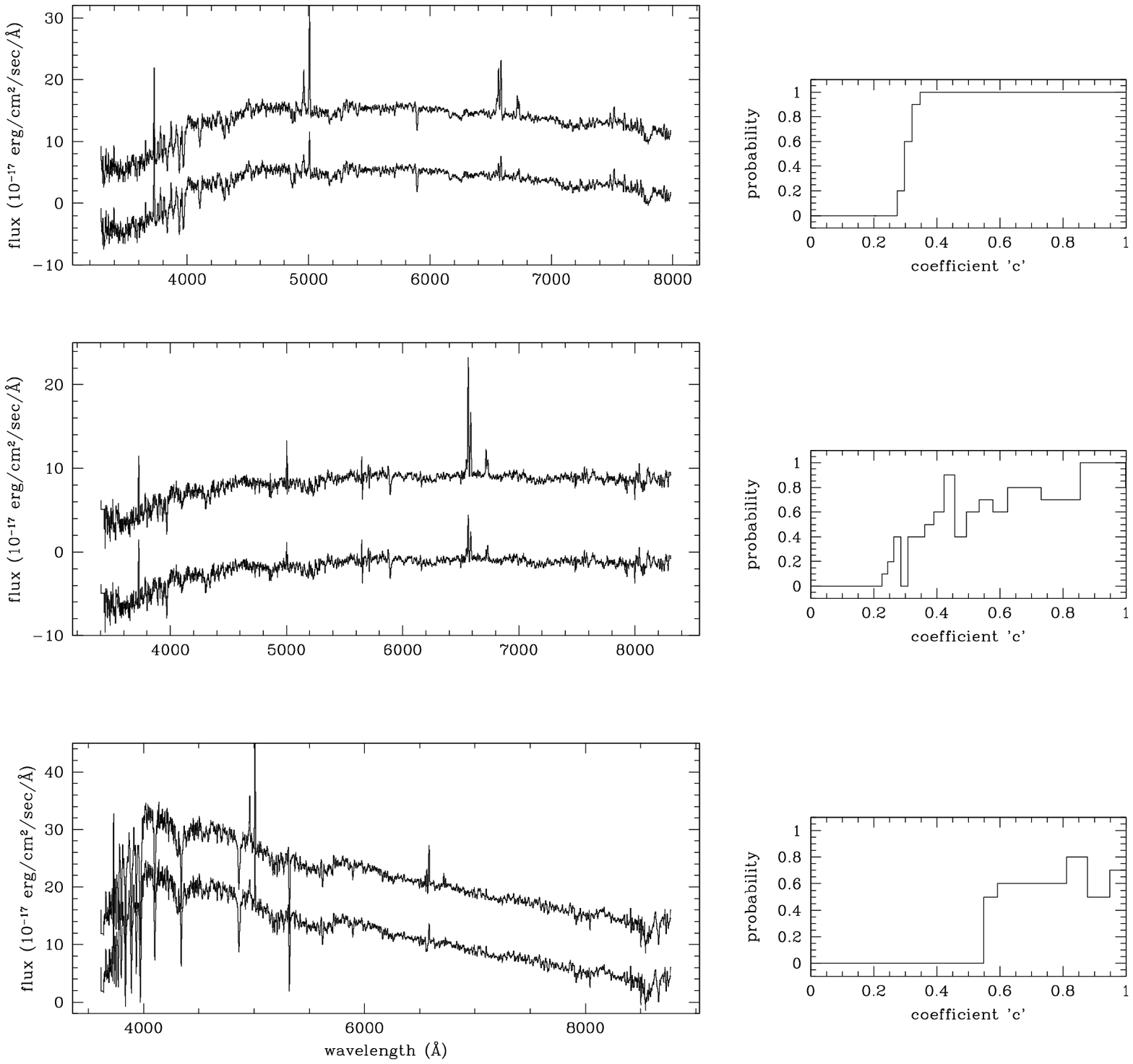}}
\caption{Three examples of our selection function measurements. The left column shows the original spectra of three AGNs and their simulated spectra after the emission lines have been scaled down by a factor of 0.4 (offset by -10). The right column plots the corresponding detection probability functions measured.}
\label{fig:sfexample}
\end{figure}

The simulation is done for each AGN in our sample, resulting in a
detection probability function as a function of $c$. For most AGN, the
detection probability function is a monotonic function of AGN
luminosity, from zero at low nuclear luminosity, to unity when the
nuclear luminosity is equal to or less than the original nuclear
luminosity, as shown in Figure~\ref{fig:sfexample}(a). For some other
AGN like that in Figure~\ref{fig:sfexample}(b), the simulation noise
overwhelms the signal and the detection probability function is not
monotonic, but fluctuates before it reaches unity. This often happens
to AGN near the limits of the selection criteria, i.e., for
narrow-line AGN, these are galaxies located close to the
AGN/starforming galaxy separation lines in the emission-line ratio
diagram; or for broad-line AGN, they have $\Ha$ FWHM close to 1,200 km
s$^{-1}$. As the nuclear emission lines become weaker, the noise
becomes more prominent and begins to affect the measured emission line
properties.

Some AGN have detection probability less than unity even when the
simulated nuclear luminosity has reached the original nuclear
luminosity, as shown in Figure~\ref{fig:sfexample}(c). One would
imagine that there is a luminosity higher than that of the original
AGN at which the detection probability would finally reach
unity. However, it is hard to correctly estimate the detection
probabilities for nuclear luminosities larger than the original
nuclear luminosity, because of the difficulty of correctly estimating
the true spectrum at higher S/N. As a result, we will simply set the
detection probabilities at nuclear luminosities larger than the
original one to be the value at the observed luminosity.

The detection probability as a function of $c$ discussed above is
calculated purely based on detectability of spectral features. In
addition, nuclear continuum in broad-line AGN contributes to the
broad-band flux. Some AGN entered our sample because the nuclear
continuum contribution pushed their $r$ band magnitudes above the
magnitude limit of $17.77$ of our sample. To quantify this effect, we
need to have some knowledge of the non-stellar nuclear continuum
contribution which we calculate from the emission line intensities.

If we assume that the emission lines we observed in our AGN spectra
are all coming from the nucleus, then they are excited by the nuclear
continuum light generated by the accretion. Therefore, the intensity
of the emission lines and their continuum must be correlated, at least
for those objects in which the continuum is unobscured. The PSF
magnitude is the closest measurement of the nuclear continuum
flux. The PSF magnitudes of most low-luminosity AGN are contaminated
by their host galaxies. However, if we can find some AGN with minimum
host galaxy contamination, we can regard their PSF luminosity
approximately as their nuclear continuum luminosity, and use them to
obtain the relationship between emission line luminosity and the
nuclear continuum luminosity. This relationship can then be applied to
other AGN.
 
We start from a list of quasars selected by their colors from the SDSS
(Richards \etal 2002) (not necessarily in our AGN sample). The fact
that they can be selected by their colors means that their host galaxy
luminosities are weak compared with their nuclear luminosities. We
apply the stellar subtraction procedure as described in Paper I to
these quasars, selecting objects with very weak or undetectable
stellar absorption line components and use these quasars to establish
the emission line luminosity vs. nuclear luminosity relationship.

\begin{figure}[t]
\centerline{\includegraphics[angle=-90, width=\hsize]{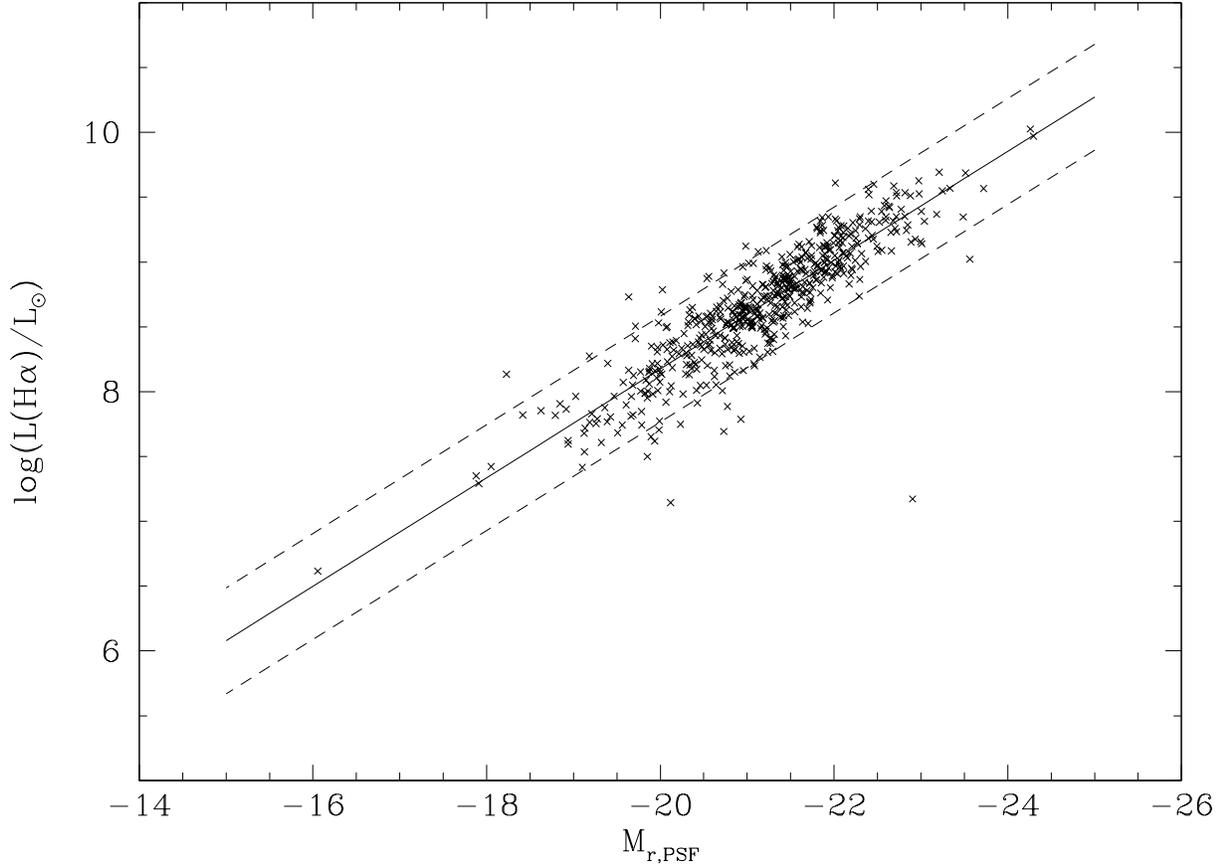}}
\caption{The $r$ band PSF magnitudes vs. the $\Ha$ luminosities for quasars with little or no stellar contamination. For these quasars, the PSF magnitudes can be approximately regarded as all coming from the nuclei. Therefore, the relationship reflected in this plot can be regarded as the relationship between the nuclear luminosity and the $\Ha$ emission line luminosity. The solid line is Equation~\ref{eq:qsohaall}. The dashed lines indicate the approximate scatter of the relationship and the differences between dashed lines and the solid line are considered as the error of Equation~\ref{eq:qsohaall}.}
\label{fig:qsoharpsf}
\end{figure}

Figure~\ref{fig:qsoharpsf} plots the $\Ha$ luminosity vs. the $r$ band
PSF absolute magnitude $M_r$ for these quasars. The relationship
between the two variables is very tight, demonstrating that the $\Ha$
luminosities for these quasars are indeed excited by their nuclear
luminosities. A linear fit to these data gives:
\begin{equation}
\log (L(\Ha)/L_{\odot})=-(0.419 \pm 0.010)\cdot M_r - (0.209 \pm 0.30)
\label{eq:qsohaall}
\end{equation}
The slope is very close to $-0.4$, which is what we expect in the
ideal case, i.e., constant $\Ha$ equivalent width. The error of the
y-intercept does not come directly from the fitting. Instead, we
directly fit the upper and lower limit of the relationship from the
plot (dashed lines in Figure~\ref{fig:qsoharpsf}) and quote 1/2 the
difference as the error. If we assume that the SEDs of high luminosity
quasars and low luminosity Seyfert 1 nuclei are similar, then
Equation~\ref{eq:qsohaall} can be used to infer the $r$ band nuclear
luminosity for our AGN from their $\Ha$ emission line strength. The
host galaxy luminosity can then be inferred by subtracting the nuclear
luminosity from the overall galaxy luminosity. As the AGN nuclear
luminosity decreases, the overall galaxy flux decreases
accordingly. The minimum nuclear luminosity for the AGN to be selected
corresponds to the overall galaxy magnitude equal to the galaxy target
criterion: r-band Petrosian $r=17.77$. This places an extra constraint
on the detection probability obtained from spectral features only.

Equation~\ref{eq:qsohaall} is obtained from a sample of quasars, with
the measured $\Ha$ flux including both broad and narrow
components. Therefore, the above description of evaluating nuclear
luminosity only works for broad-line AGN. For narrow-line AGN in our
sample, fortunately, the non-stellar continuum contribution is
negligible compared to the host galaxy luminosity (Kauffmann \etal
2003). Therefore, for narrow-line AGN, we simply skip the above
analysis and keep the detection probability obtained from the spectral
features only.

\section{Measuring the Luminosity Function}

%Assume the luminosity function is parameterized as $N_p$ steps:
%$$ \Phi(L) = \phi_k, \,\,\,\,\,L_k- \frac{\Delta L}{2} < L <  L_k+ \frac{\Delta L}{2}, \,\,\,\,\, k = 1, ......, N_p $$
%So the likelihood $ {\cal L}$ is:
%$$\ln {\cal L} = \sum_{i=1}^N W(L_i -L_k) \ln \phi_k - \sum_{i=1}^N \ln \left[ \sum_{j=1}^{N_p}\phi_j \Delta L \; H(L_j - L_{min,\, i})  \right] + {\rm const} $$
%
%
%$$ {\cal L} = \prod_i {\cal L}_i = \prod_i \frac {p_i(L_i) \Phi (L_i) dL_i}{\int \Phi(L) p_i(L)dL}$$

The AGN luminosity function $\Phi(L)$ is defined as the mean number of
AGN per unit volume, per unit luminosity. In making the plot of the
luminosity function without covering several magnitudes of luminosity,
we also define the auxiliary quantity $\hat\Phi(L)$: the distribution
of AGN per unit volume per log luminosity, given by:
\begin{equation}
\hat\Phi(L) = -\frac{d\Psi(L)} {d(\log_{10} L)} = {{1} \over {\log_{10} e }} \, L\Phi(L),
\end{equation}
where $\Psi(L)$ is the cumulative luminosity function, and is related
to $\Phi(L)$ by $\Phi(L)={d\Psi(L)}/{dL}$. $\hat\Phi(L)$ is widely
used when making luminosity function plots. Throughout this paper, the
luminosity functions are all plotted in this form.

The luminosity function can be measured via various methods. In this
paper, we will use the non-parametric method which does not require
any assumption about the luminosity function shape.

The likelihood that each AGN $i$, with detection probability function
$p_i(L)$ (described in $\S2$) have luminosity L, given its redshift
and host galaxy properties, is:
\begin{equation}
{\cal L}_i = \frac {p_i(L) \Phi (L) dL}{\int \Phi(L) p_i(L)dL}.
\end{equation}

The maximum likelihood solution $\Phi(L)$ is the function that
maximizes the overall likelihood
\begin{equation}
{\cal L} = \prod_i {\cal L}_i = \prod_i \frac {p_i(L_i) \Phi (L_i) dL_i}{\int \Phi(L) p_i(L)dL}.
\end{equation}
%$$\ln {\cal L} = \sum_{i=1}^N W(L_i -L_k) \ln \phi_k - \sum_{i=1}^N \ln \left[ \sum_{j=1}^{N_p}\phi_j \Delta L \; H(L_j - L_{min,\, i})  \right] + {\rm const} $$
$\Phi(L)$ is obtained by minimizing ${\cal S}\equiv -2 \ln {\cal
L}$. In this paper, we use this method to obtain non-parametric,
linear-interpolated stepwise (Koranyi \& Strauss 1997) luminosity
functions. The detailed derivation can be found in Efstathiou, Ellis
\& Peterson (1988), Koranyi \& Strauss (1997) and Blanton (2000).

The luminosity function $\Phi (L)$ derived above is not
normalized. The normalization is done by directly measuring the number
density of AGN, expressed as:
\begin{equation}
n = \frac{1}{V} \sum_i\frac {\int_0^\infty \Phi (L) dL}{\int_0^\infty \Phi(L) p_i(L)dL},
\label{eq:norm}
\end{equation}
where V is the volume spanned by the galaxies in the sample and the
sum is over all AGN included in this luminosity function
calculation. The error associated with the normalization is:
\begin{equation}
n_{err} = \frac{1}{V} \left (\sum_i \left ( \frac {\int_0^\infty \Phi (L) dL}{\int_0^\infty \Phi(L) p_i(L)dL}\right )^2\right )^{1/2}.
\label{eq:normer}
\end{equation}

The final normalized luminosity function is then $\phi (L) = n\cdot
\Phi (L)$.

\section{Luminosity Function Results}
\subsection{Overall $\Ha$ Luminosity Function}
\begin{figure}[t]
\centerline{\includegraphics[angle=-90, width=\hsize]{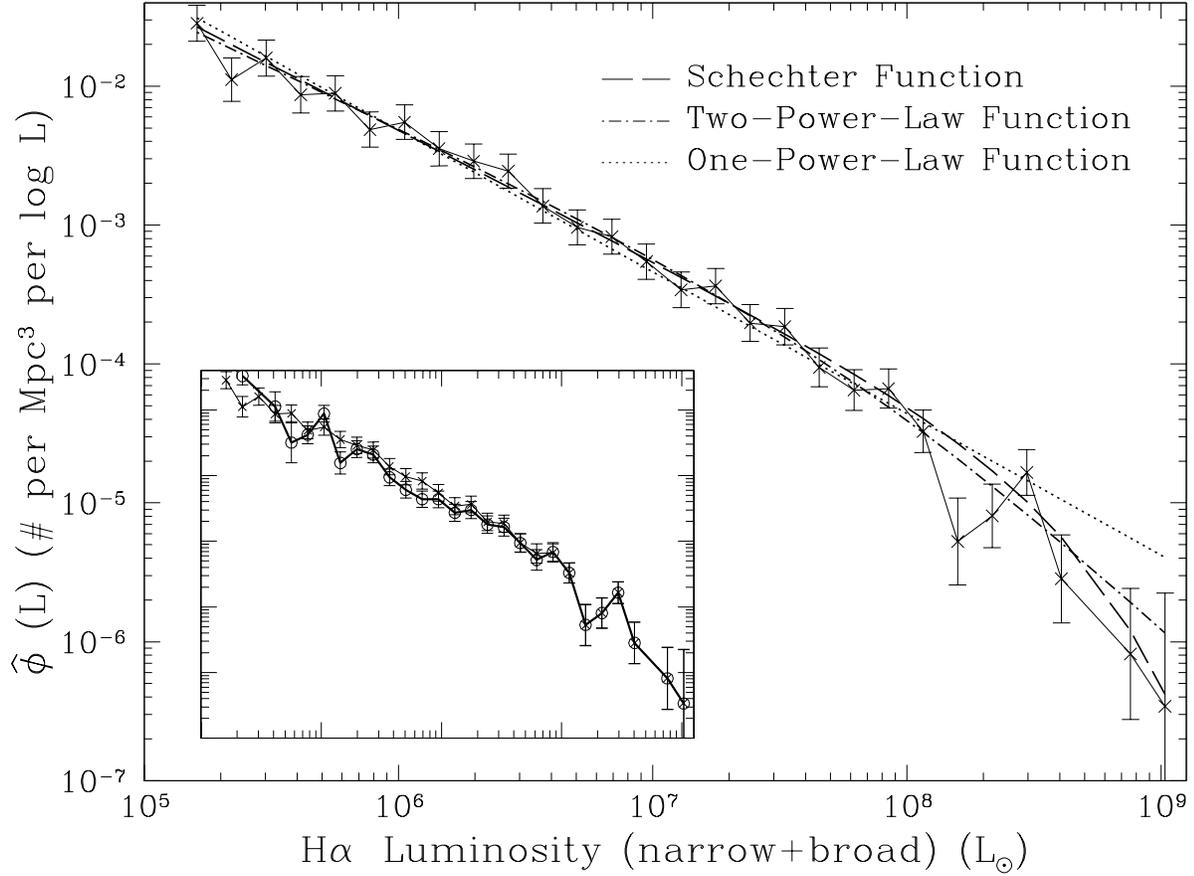}}
\caption{The overall $\Ha$ luminosity function for Seyferts, including both Seyfert 1s and Seyfert 2s. We fit the luminosity function with Schechter function (Equation~\ref{eq:sch}), Two-Power-Law function (Equation~\ref{eq:2p}) and One-Power-Law function (Equation~\ref{eq:1p}). The best fitting parameter and corresponding $\chi^2$ are listed in Table~\ref{tab:fit}. The inserted plot shows the Seyfert 1 $\Ha$ luminosity function (thick line with open circles) compared with the overall $\Ha$ luminosity function.}
\label{fig:lfhaall}
\end{figure}

The overall Seyfert luminosity function can be obtained by combining
the Seyfert 1 and Seyfert 2 samples. Here we choose to use the
narrow-line AGN sample selected via Kewley's criteria, because it is
not significantly contaminated by star formation activity. The overall
AGN sample includes about 4400 AGN: 1317 Seyfert 1s and 3074 Seyfert
2s in the redshift range $0<z<0.33$. Since objects at higher redshifts
have lower S/N and therefore higher uncertainties in the evaluation of
the selection function, we limit our measurement of the luminosity
function to $0<z<0.15$; this excludes about one third of the
sample. The selection functions for these Seyferts takes into account
whether an object is classified either as Seyfert 1 or Seyfert 2 at
each luminosity, which includes the cases when the classification
changes between Seyfert 1 and Seyfert 2 with changed emission line
strength. Using the methodology discussed above, we obtain the overall
$\Ha$ luminosity function and plot it in Figure~\ref{fig:lfhaall}.

The Seyfert $\Ha$ luminosity function covers a luminosity range of a
factor of $10^4$. We fit the luminosity function with a two-power-law
formula as used in the 2dF quasar luminosity function by Croom \etal
(2004):
\begin{equation}
\phi(L)=\frac{\phi^*(L_*)/L_*}{\left(\frac{L}{L_*}\right)^\alpha+\left(\frac{L}{L_*}\right)^\beta},
\label{eq:2p}
\end{equation}
where $\phi^*(L_*)$, $L_*$, $\alpha$ and $\beta$ are free
parameters. The best fit results and the $\chi^2$ of the fit are
listed in Table~\ref{tab:fit}.  We can also fit the curve to the
Schechter (1976) function:
\begin{equation}
\phi(L)=\frac{\phi^*(L_*)}{L_*}\left(\frac{L}{L_*}\right)^\alpha \exp \left(-\frac{L}{L_*}\right),
\label{eq:sch}
\end{equation}
and the best fit $\phi^*(L_*)$, $L_*$, $\alpha$ are listed in
Table~\ref{tab:fit} as well. Noticing that the luminosity function
shape is also close to a single power-law, we fit it to a single
power-law form:
\begin{equation}
\phi(L)=\frac{\phi^*(L_*)}{L_*}\left(\frac{L}{L_*}\right)^\alpha
\label{eq:1p}
\end{equation}
Again, the best fit $\phi^*(L_*)$, $L_*$, $\alpha$ are listed in
Table~\ref{tab:fit} (All following luminosity functions are fit with
the three models and the results are all listed in
Table~\ref{tab:fit}).  The three best-fit functions are also plotted
in Figure~\ref{fig:lfhaall}. The shape of the AGN overall $\Ha$
luminosity function is closer to a Schechter function or a two-power
law shape, even though it does not differ much from a single power-law
(see $\chi^2$ values in Table~\ref{tab:fit} for comparison).

The AGN number density can be estimated by integrating the AGN
luminosity function over the luminosity range it covers. Using the
best-fit Schechter function and within the $\Ha$ luminosity range of
$10^5 L_\odot$ to $10^9 L_\odot$, we obtain the AGN number density as
$0.018 \ {\rm Mpc}^{-3}$. To compare this number with the galaxy
number density, we integrate the galaxy luminosity function obtained
from the SDSS (Blanton \etal 2003b) over an absolute magnitude range
of $-14<M_r<-24$ and obtain the galaxy number density to be $0.094
{\rm Mpc}^{-3}$ in the same cosmology. Thus active galaxies in this
luminosity range comprise about 19\% of all galaxies (Ho \etal 1997b;
Miller \etal 2003; Brinchmann \etal 2004).

Similarly, we can also obtain the luminosity density by integrating
$L\phi(L)$ over the entire luminosity range. Since the overall $\Ha$
luminosity function is a mix of broad $\Ha$ plus narrow $\Ha$ for
Seyfert 1s and narrow $\Ha$ only for Seyfert 2s, and since narrow-line
AGN contribute little to the $r$ band luminosity, it is appropriate to
apply the integration to just the broad-line AGN $\Ha$ luminosity
function. This LF is derived in a similar way as the overall $\Ha$
luminosity function, and the result is shown in the inserted plot of
Figure~\ref{fig:lfhaall} (thick line with open circles). The $\Ha$
luminosity function for all Seyferts is over-plotted (thin line with
crosses) for comparison. Seyfert 1 $\Ha$ luminosity is very close to
the overall Seyfert $\Ha$ luminosity even at the low luminosity
range. The AGN $\Ha$ luminosity density integrated over this
luminosity function is $1.54\times 10^4 L_\odot {\rm Mpc}^{-3}$. Using
the relationship between $\Ha$ luminosity and AGN $r$-band continuum
luminosity (Equation~\ref{eq:qsohaall}), we obtain the AGN $r$-band
luminosity density as $1.11\times 10^6 L_\odot {\rm
Mpc}^{-3}$. Comparing with the galaxy luminosity density obtained from
Blanton \etal (2003b): $1.84\times 10^8 L_\odot {\rm Mpc}^{-3}$, we
find that the nuclear activity contributes about $6\times 10^{-3}$ of
the total light of galaxies in the $r$-band.

\subsection{Seyfert 1 vs. Seyfert 2 luminosity function}
Our AGN sample is also sufficiently large to evaluate the narrow- and
broad-line AGN luminosity functions separately. The comparison of the
two can be used to demonstrate the relative number ratio of Seyfert 1s
and Seyfert 2s. For this task, the overall $\Ha$ luminosity is no
longer an appropriate parameter for the luminosity function
measurements. According to the unification model, the observed overall
$\Ha$ luminosity in Seyfert 1s includes both that emitted from the
broad-line region and that from the narrow-line region, but the broad
component is obscured in Seyfert 2s and only the narrow component is
observed. Therefore, the same measured overall $\Ha$ luminosities for
broad-line and narrow-line AGN could correspond to different nuclear
continuum luminosities and therefore the differences of the overall
$\Ha$ luminosity function of the two types of AGN do not necessarily
reflect the real differences of the number density of Seyfert 1s and
Seyfert 2s (see the inserted plot of Figure~\ref{fig:lfhaall}, Seyfert
1 luminosity function dominates even at the low luminosity range). A
good luminosity function parameter would be emission line luminosities
that are isotropic to Seyfert 1s and Seyfert 2s. In our study, the
$\Ha$ narrow-line component, [OIII] and [OII] luminosities are good
candidates.

\begin{figure}[t]
\centerline{\includegraphics[angle=-90, width=\hsize]{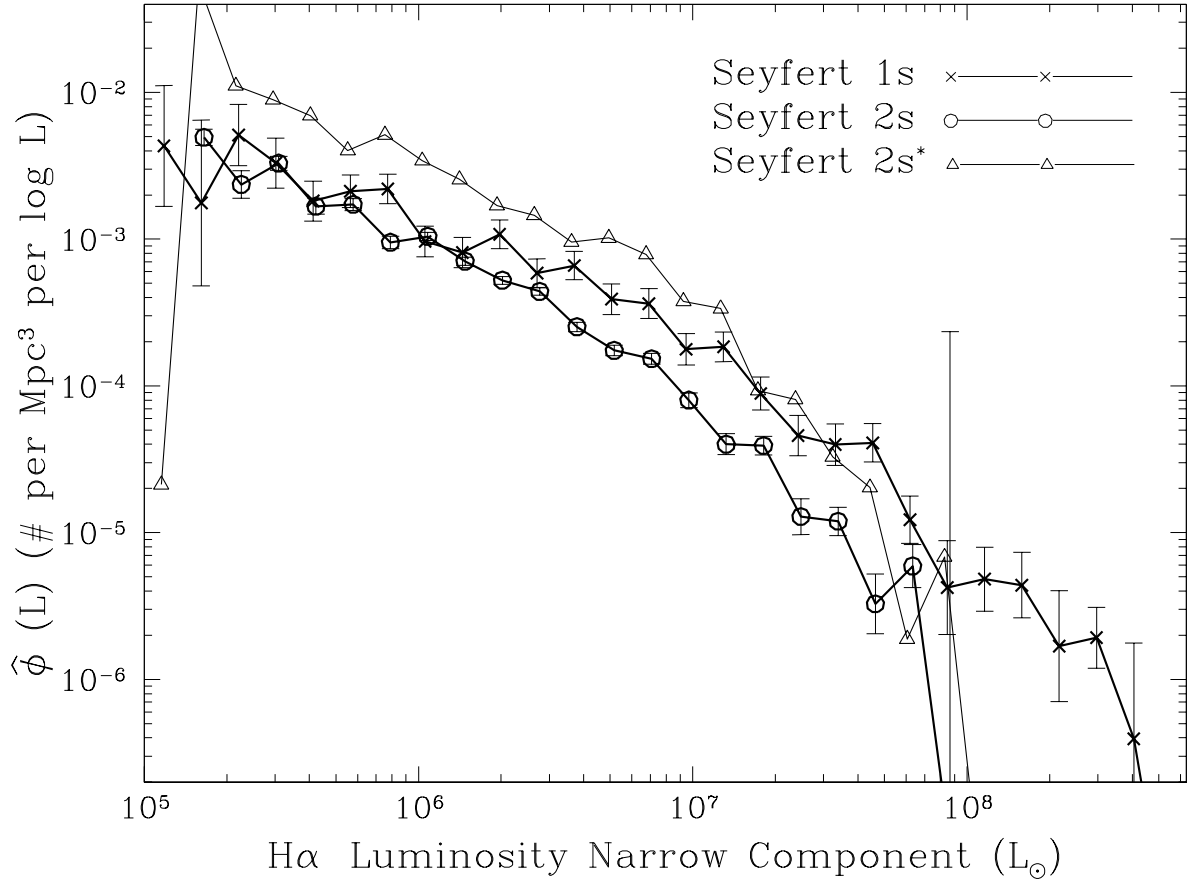}}
\caption{The narrow-line component $\Ha$ luminosity functions for Seyfert 2s (open circles, selected via Kewley's criterion), Seyfert 1s (crosses) and Seyfert $2^*$s (triangles, selected via Kauffmann's criterion) separately. The Seyfert 1 and Seyfert 2 luminosity functions are about the same at low luminosity. But at high luminosity, the Seyfert 2 luminosity function drops off more quickly than does the Seyfert 1 luminosity function. The Seyfert $2^*$s luminosity function is significantly larger than that of Seyfert 2s. To avoid crowding the plot, we didn't show errorbars of the Seyfert $2^*$ luminosity function, which are comparable to the errorbars of the other two luminosity functions. The same applies to the following three plots.}
\label{fig:lf3han}
\end{figure}

\begin{figure}[t]
\centerline{\includegraphics[angle=-90, width=\hsize]{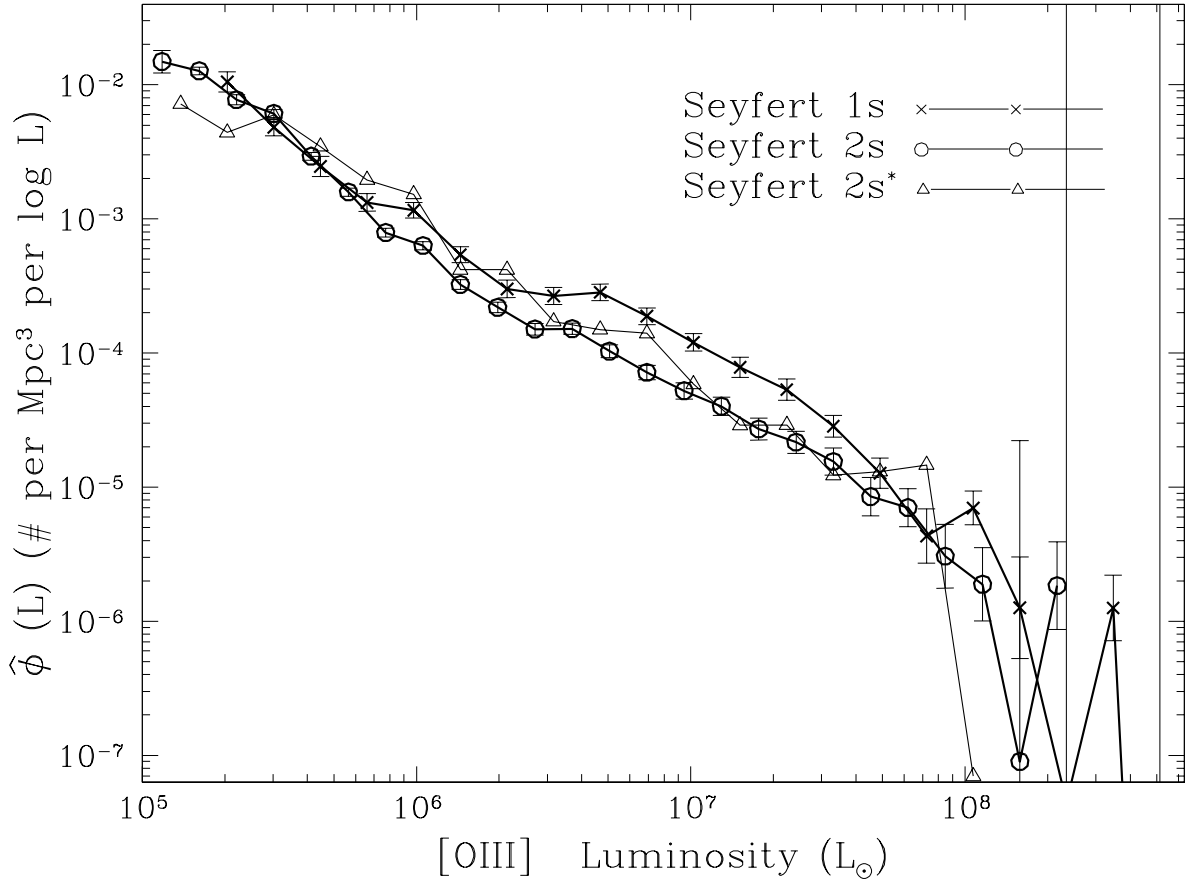}}
\caption{The [OIII] luminosity functions for Seyfert 2s (open circles), Seyfert 1s (crosses) and Seyfert $2^*$s (triangles) separately.  The comparison of the Seyfert 1 and Seyfert 2 luminosity functions shows us that the number ratio of Seyfert 1s and Seyfert 2s is a function of luminosity. At low luminosity, the number density of Seyfert 1s and Seyfert 2s are about the same, but at high luminosity, Seyfert 1s gradually outnumber Seyfert 2s. The fact that the Seyfert $2^*$ luminosity function is comparable to that of Seyfert 2s demonstrates that we are not missing much AGN activity in the Seyfert 2s sample. The comparison of the three confirms our conclusion of the relative number ratio of Seyfert 1s and Seyfert 2s.}
\label{fig:lf3o3}
\end{figure}

\begin{figure}[t]
\centerline{\includegraphics[angle=-90, width=\hsize]{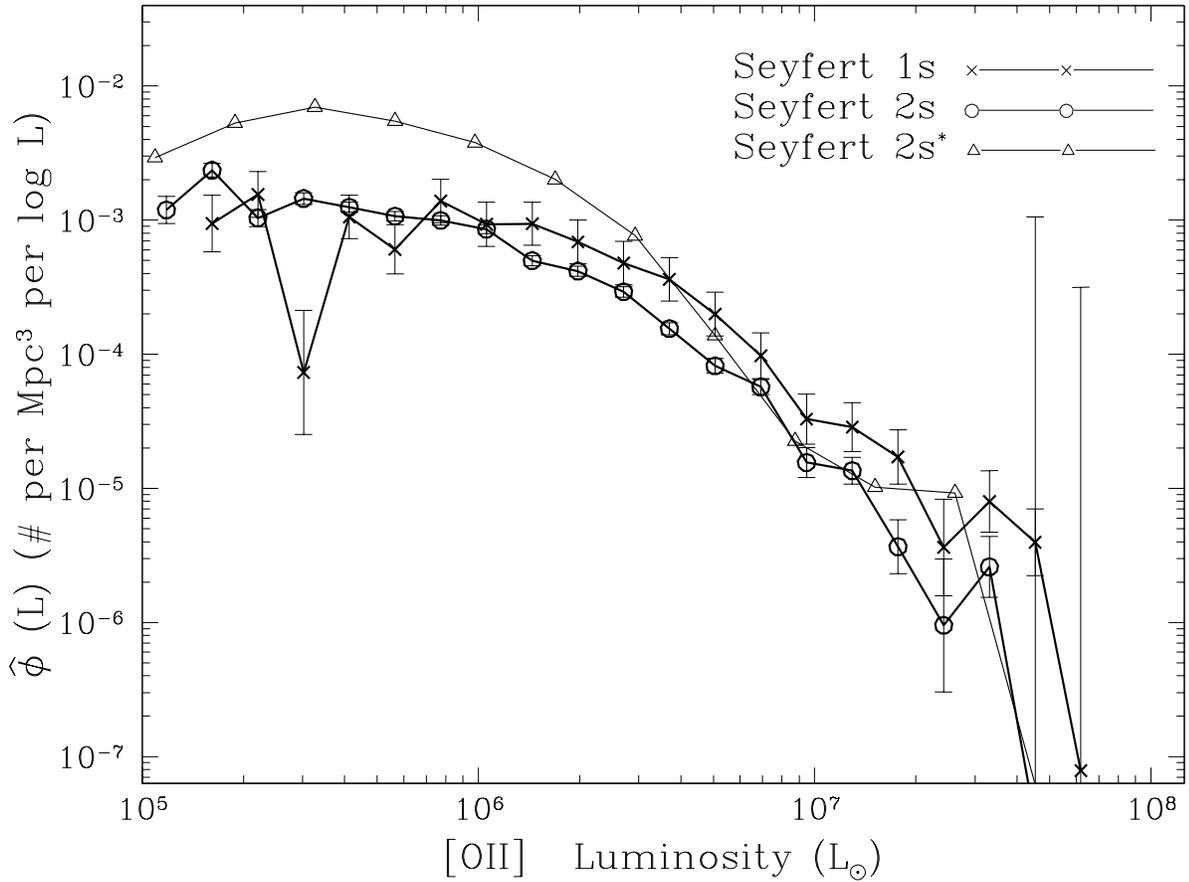}}
\caption{The [OII] luminosity functions for Seyfert 2s (open circles), Seyfert 1s (crosses) and Seyfert $2^*$s (triangles) separately.  The comparison of the Seyfert 1 and 2 luminosity functions shows the same pattern as of Figure~\ref{fig:lf3o3}. Again, the Seyfert $2^*$ luminosity function is significantly larger than that of Seyfert 2s because of star formation contamination.}
\label{fig:lf3o2}
\end{figure}

Since we have two narrow-line AGN samples: one selected via Kewley's
criteria, which are called Seyfert 2 sample; and another one selected
via Kauffmann's criterion, which we call Seyfert 2$^*$ sample, we will
first compare the Seyfert 1 luminosity function with the Seyfert 2
luminosity function, then with the Seyfert 2$^*$ luminosity
function. In obtaining the luminosity functions, the selection
functions of objects in each sample are evaluated with the selection
criteria only corresponding to this sample in question. Therefore,
unlike the case of the total Seyfert sample, the detection probability
is calculated for Seyfert 1s and Seyfert 2s
separately. Figure~\ref{fig:lf3han}, ~\ref{fig:lf3o3} and
~\ref{fig:lf3o2} plot the $\Ha$ narrow-line component, [OIII] and
[OII] emission-line luminosity functions for Seyfert 1s (crosses) and
Seyfert 2s (open circles) respectively. The differences between the
luminosity functions of the two types of AGN all show the same
pattern: they are comparable at low luminosity, but at high
luminosities, the Seyfert 2 luminosity function drops off more quickly
than does that of Seyfert 1s. This demonstrates that the number ratio
of Seyfert 1s and Seyfert 2s is a function of luminosity: at
low-luminosity, the number density of the two types of AGN are about
the same, but at high luminosity, Seyfert 1s outnumber Seyfert 2s.

One big concern of our argument of the relative number ratio of
Seyfert 1s and Seyfert 2s comes from the way we define our Seyfert
sample. In particular, in Seyfert 2$^*$ sample, which is selected via
the Kauffmann's criterion, we have about 10,700 narrow-line AGN. This
is almost 3 times larger than the Seyfert 2 sample selected via the
Kewley's criteria. Our Seyfert 1 sample however is not affected by
this change of criteria. Therefore selecting AGN via different
criteria could dramatically change our conclusions about the number
ratio of Seyfert 1s and Seyfert 2s.

To test this, we further measure the [OIII], $\Ha$ and [OII]
luminosity function for the Seyfert 2$^*$ sample
(Figure~\ref{fig:lf3han}, ~\ref{fig:lf3o3} and ~\ref{fig:lf3o2} open
triangles), which is selected via Kauffmann's criterion. Because of
the large number of objects included in the Seyfert $2^*$ sample, and
because our determination of the selection function typically requires
several hundred Monte-Carlo simulations per object, we are unable to
measure the probability function for each galaxy as we did for our
Seyfert 1 and Seyfert 2 samples. Instead we randomly select 1/8 of the
sample and measure their selection functions, being sure to multiply
the overall normalization by 8.

Except for the high luminosity end, the $\Ha$ and [OII] luminosity
function of the Seyfert 2$^*$ sample is larger than that of Seyfert 2s
by a factor of about 4. The Seyfert 2$^*$ [OIII] luminosity function,
on the other hand, is only slightly larger than that of Seyfert 2s at
intermediate luminosity. This is consistent with results of Kauffmann
\etal (2003). In figure 2 of the paper, it can be seen that most of
the Seyfert 2s that are included by Kauffmann's criterion but not by
Kewley's criterion are those that have intermediate [OIII]
luminosities. The results shown in Figure~\ref{fig:lf3han},
~\ref{fig:lf3o3} and ~\ref{fig:lf3o2} agree with the analysis
described in Paper I and Section 1: Seyfert $2^*$s selected via
Kauffmann's criterion tend to include all AGN activity, but suffer
from contamination from star formation. The only exception is the
[OIII] emission line, as the [OIII] contamination from star formation
is relatively weak. Kewley's criterion on the other hand only selects
those galaxies that are dominated by AGN activity. Therefore, due to
the contamination, the $\Ha$ and [OII] luminosity functions of the
Seyfert 2$^*$ sample include a star formation component and are
significantly higher than those of Seyfert 2 sample, while the [OIII]
luminosity function more or less reflects the real AGN
distribution. When comparing the Seyfert 2$^*$ [OIII] luminosity
function with that of Seyfert 1s, we obtain the same conclusion for
the relative ratio of Seyfert 1s and Seyfert 2s as before: they are
comparable at low luminosity, but Seyfert 1s outnumber Seyfert 2s at
high luminosity.
 
\begin{figure}[t]
\centerline{\includegraphics[angle=-90, width=\hsize]{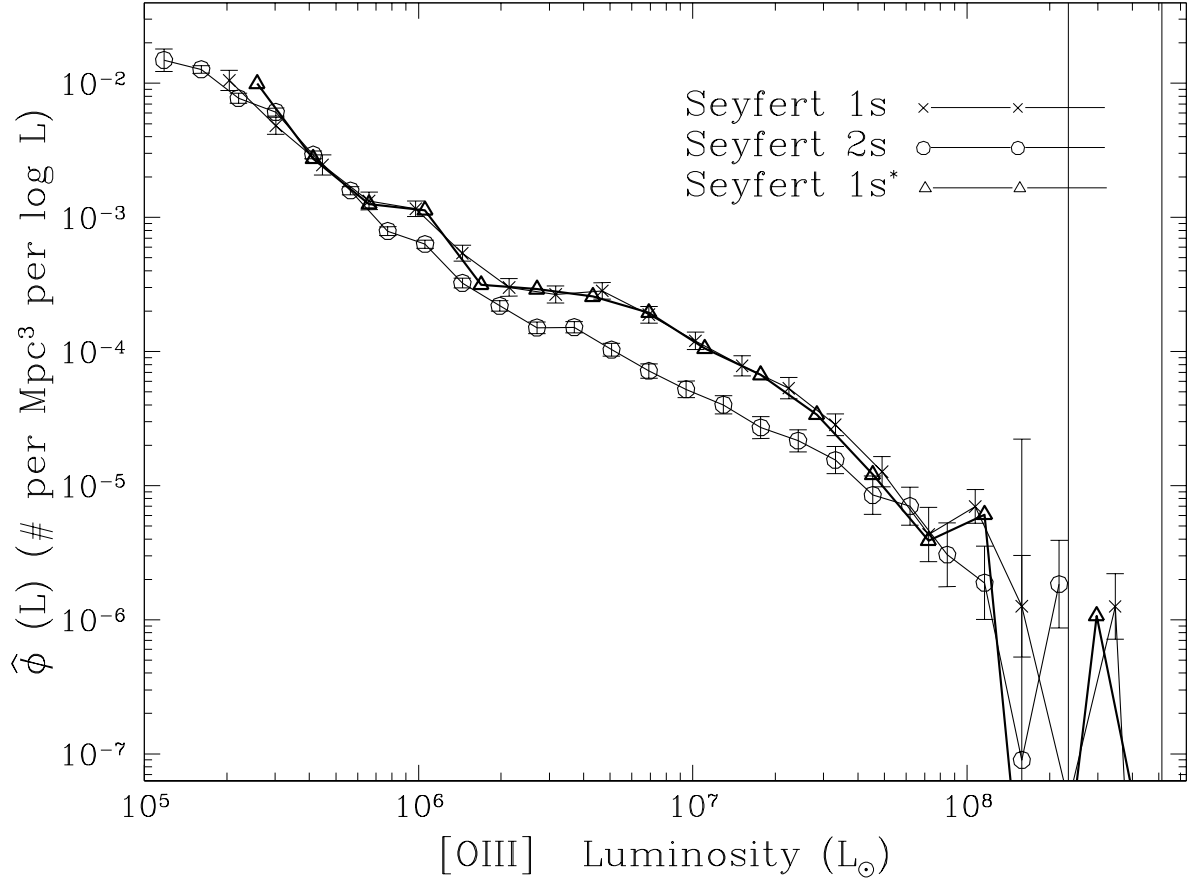}}
\caption{The [OIII] luminosity functions for Seyfert 2s (open circle), Seyfert 1s (crosses) and Seyfert 1s (triangles) whose narrow components also satisfy the narrow-line AGN selection criteria (noted as Seyfert 1s$^*$). There is little differences between the luminosity functions for Seyfert 1s and Seyfert 1s$^*$ and the comparison of the luminosity functions of Seyfert 2s and Seyfert 1s$^*$ reflects the number ratio of the two types of AGNs free of the selection criteria uncertainties. } 
\label{fig:lf3o3b}
\end{figure}

A further test for the robustness of our luminosity function results
can be applied by constraining the Seyfert 1 sample to include only
those objects whose narrow $\Ha$ component also satisfy the Seyfert 2
selection criteria as well (see Figure 10 in Paper I). We refer to such
objects as Seyfert 1$^*$s. The probability function for the sample is
obtained by requiring the simulated AGN to satisfy both the Seyfert 1
and Seyfert 2 criteria. Figure~\ref{fig:lf3o3b} shows the comparison
of the [OIII] luminosity functions obtained from the Seyfert 1,
Seyfert 2 and Seyfert 1$^*$ samples. The two broad-line AGN sample
luminosity functions are almost the same; in particular, the
comparison of the Seyfert 2 and Seyfert 1$^*$ luminosity functions
shows the same behavior described above. Thus our conclusion of the
relative number ratio of Seyfert 1s and Seyfert 2s is not sensitive to
the selection criterion uncertainties.

\section{The Relationship between AGN and Host Galaxy Luminosity}
\begin{figure}[t]
\centerline{\includegraphics[width=0.4\hsize, angle=-90]{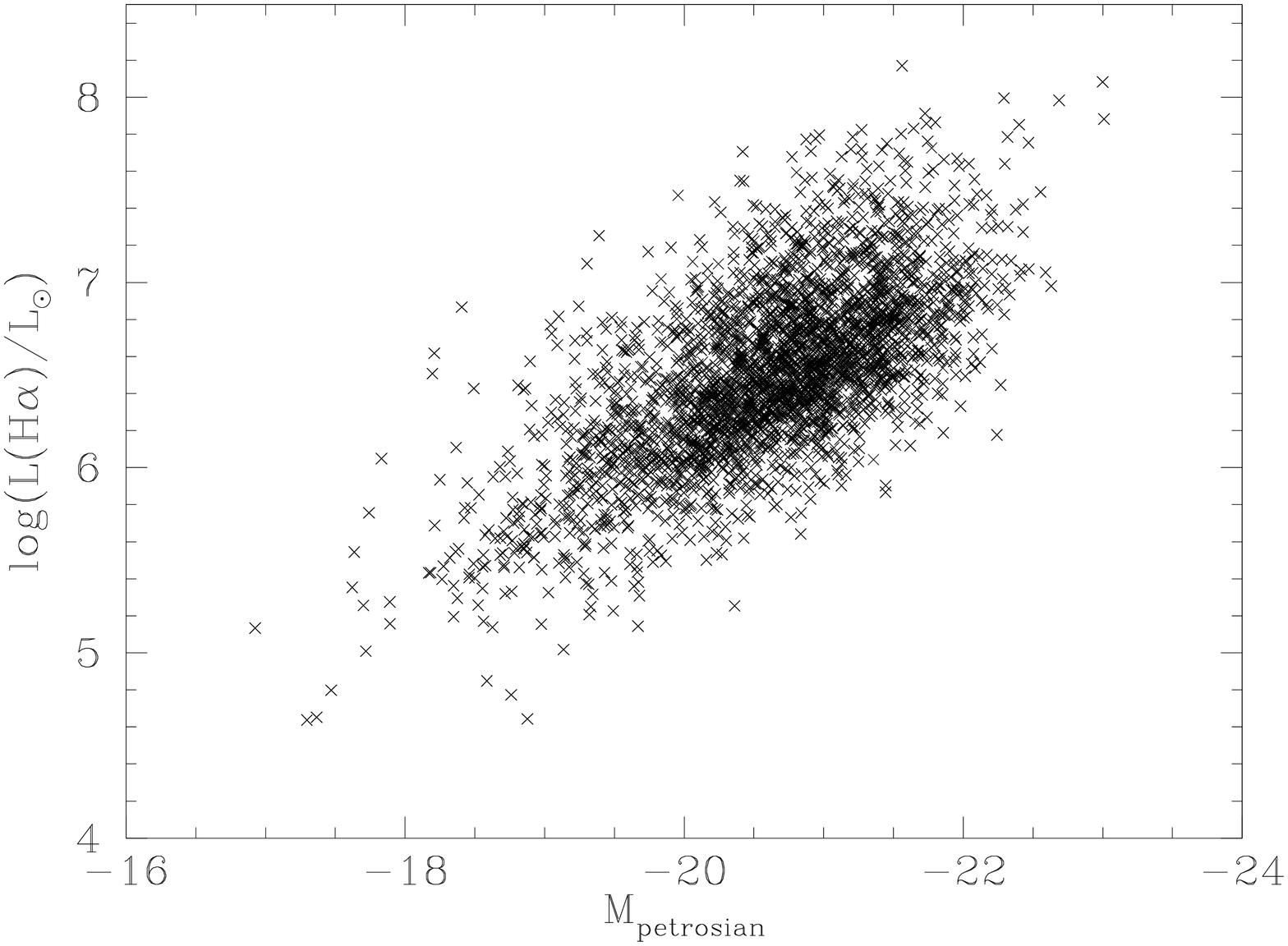}\includegraphics[width=0.4\hsize, angle=-90]{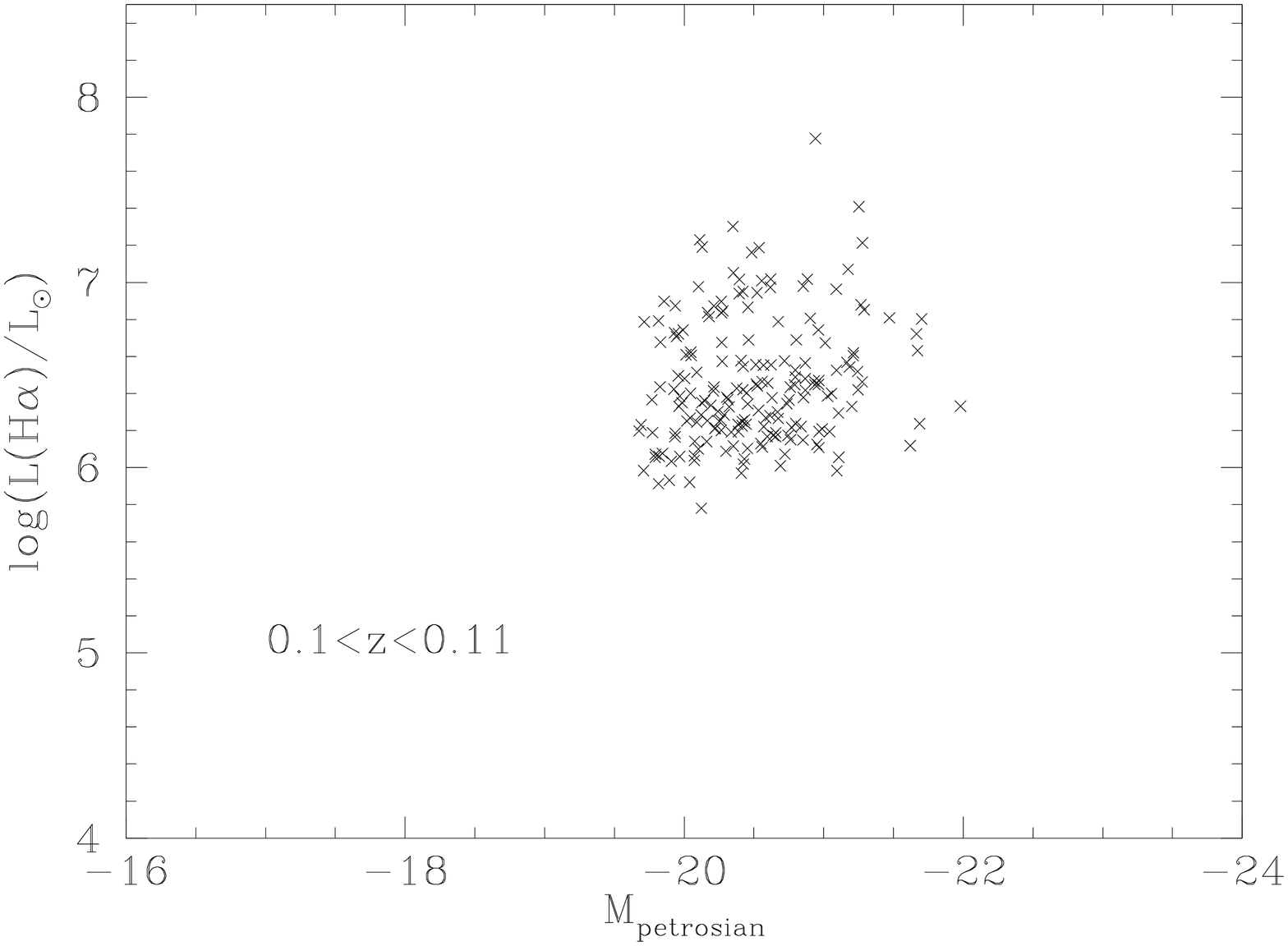}}
\caption{The $r$ band Petrosian magnitudes vs. the $\Ha$ luminosities for the narrow-line AGN sample. It appears that there is a correlation between the two luminosities. But the apparent correlation is very likely to be due to selection effects. When we choose a subsample with small redshift range, so that the selection effects are not significant (shown in the right figure), we find little correlation between the two luminosities.}
\label{fig:hastrmag}
\end{figure}

The selection functions we have obtained are based on the assumption
that the nuclear luminosity is independent of the host galaxy
luminosity. We can check this assumption by plotting the $\Ha$
emission line luminosity against the host galaxy Petrosian luminosity
for our AGN sample. Figure~\ref{fig:hastrmag}(a) shows the plot for
Seyfert 2s. There seems to be a strong correlation between the two
variables: brighter host galaxies appear to have stronger $\Ha$
emission lines. But the apparent lack of galaxies at the lower-right
corner of the plot is due to a selection effect: if the host galaxy is
too luminous relative to the nuclear emission line strength, the host
galaxy will overwhelm the active nuclear features and the object
cannot be recognized as an AGN. The lack of objects on the upper-left
corner, on the other hand, simply reflects the fact that AGN nuclear
luminosities have upper limits: their Eddington luminosities. If we
plot these AGN in small redshift ranges over which the selection
effect is weak, \eg as shown in Figure~\ref{fig:hastrmag}(b), we see
little correlation between the two variables. We now carry out a more
quantitative test, which involves using our luminosity function
results.

Imagine that our assumption is wrong, and for Seyferts, the nuclear
luminosity is generally stronger in a more luminous host galaxy. Then
based on the wrong assumption, the obtained detection probability
function will be underestimated at low luminosity and overestimated at
higher luminosity. Therefore, the resultant luminosity function will
reflect a lower density of low-luminosity objects and higher density
of high-luminosity objects than there should be. If we use this
incorrect luminosity function to evaluate the expected luminosity
distribution of a subgroup of objects with, for example, low redshift
or low host galaxy luminosity, the resultant distribution will be
systematically shifted to higher luminosity than the observed
distribution. Similarly, for a subgroup of objects with high redshift
or high host galaxy luminosity, the calculated luminosity distribution
will be shifted to lower luminosity than the observed
distribution. Only if our assumption is correct will the expected
luminosity distribution match up with the observed luminosity
distribution for any subgroup of our sample. Based on these
considerations, we apply the following tests to each of our AGN
samples.

\begin{figure}[t]
\centerline{\includegraphics[angle=-90, width=\hsize]{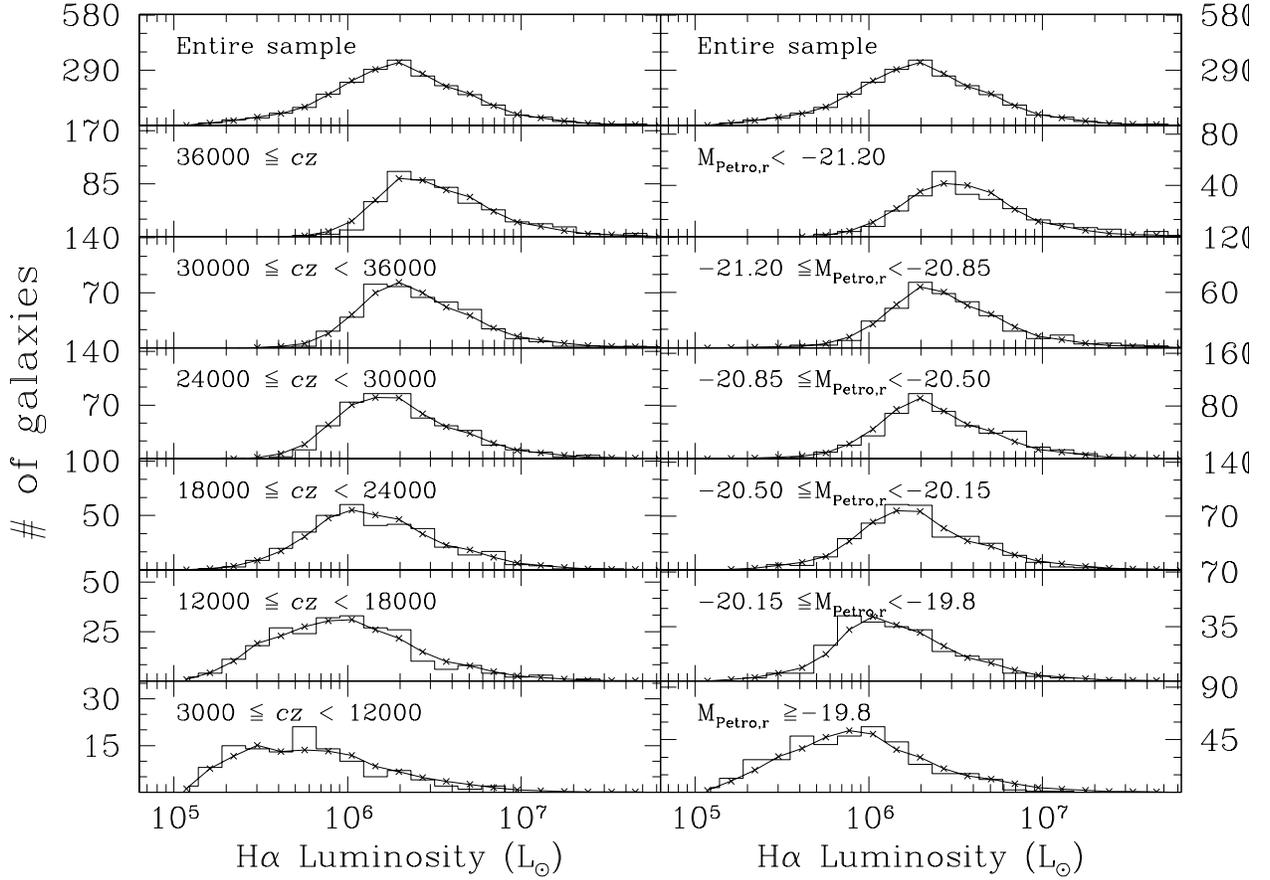}}
\caption{ The observed luminosity distribution for Seyfert 2s (histogram) compared with the expected luminosity distribution (solid line) calculated from the Seyfert 2 $\Ha$ luminosity function (Figure~\ref{fig:lf3han}) for different subgroups of the Seyfert 2s sample. {\it Left}: the entire Seyfert 2 sample (shown in the top panel) is grouped by their redshifts. {\it Right}: the entire Seyfert 2s sample is grouped by their absolute $r$ band Petrosian magnitudes. The match up between the observed luminosity distribution and the expected luminosity distribution in each subgroup demonstrates that our luminosity function results are correct and the assumption that the host galaxy luminosity is independent of the nuclear luminosity is reasonable.}
\label{fig:lfckhann}
\end{figure}

We divide our Seyfert samples (either a Seyfert 1, Seyfert 2 or
Seyfert $2^*$ sample) into several subgroups by their redshifts. For
each subgroup, we use our Seyfert luminosity function results to
evaluate the AGN luminosity distribution we expect to observe (i.e.,
the number of Seyferts in the samples as a function of redshift),
which can be written as:
\begin{equation}
F_{expect}(L)\Delta L = \sum_i \frac{p_i(L) \Phi (L) \Delta L}{\int_0^\infty \Phi(L^\prime) p_i(L^\prime)dL^\prime} 
\end{equation}
where the sum is over all Seyferts in a subsample. We compare
$F_{expect}(L)$ with the observed number of Seyferts in this subsample
between $L$ and $L+\Delta L$. Figure~\ref{fig:lfckhann}(a)
demonstrates the comparison using the Seyfert 2 $\Ha$ luminosity
function. The observed and expected luminosity distributions agree
very well with each other. It is not surprising that the two match
perfectly for the entire Seyfert 2 sample, since the luminosity
function is measured from this sample. But the fact that they agree in
each redshift bin demonstrates that our luminosity function reflects
the true Seyfert 2 distribution. In addition, this demonstrates that
our assumption of the independence of AGN and host galaxy luminosity
is reasonable.

We next run a similar test by dividing the Seyfert 2 sample via host
galaxy luminosity. By doing so, we can directly examine the
relationship between nuclear luminosities and host galaxy
luminosities.  Figure~\ref{fig:lfckhann}(b) shows the result of this
test for the Seyfert 2 $\Ha$ luminosity function. Higher luminosity
Seyfert 2 galaxies have more luminous $\Ha$ emission lines (as they
tend to be at higher redshift on average), as we saw in
Figure~\ref{fig:hastrmag}(a). However, in each subgroup, the observed
and expected luminosity distributions match up impressively well,
which gives us a quantitative confirmation that the nuclear
luminosities for Seyfert 2s are indeed not strongly correlated with
the host galaxy luminosities.

\begin{figure}[t]
\centerline{\includegraphics[angle=-90, width=\hsize]{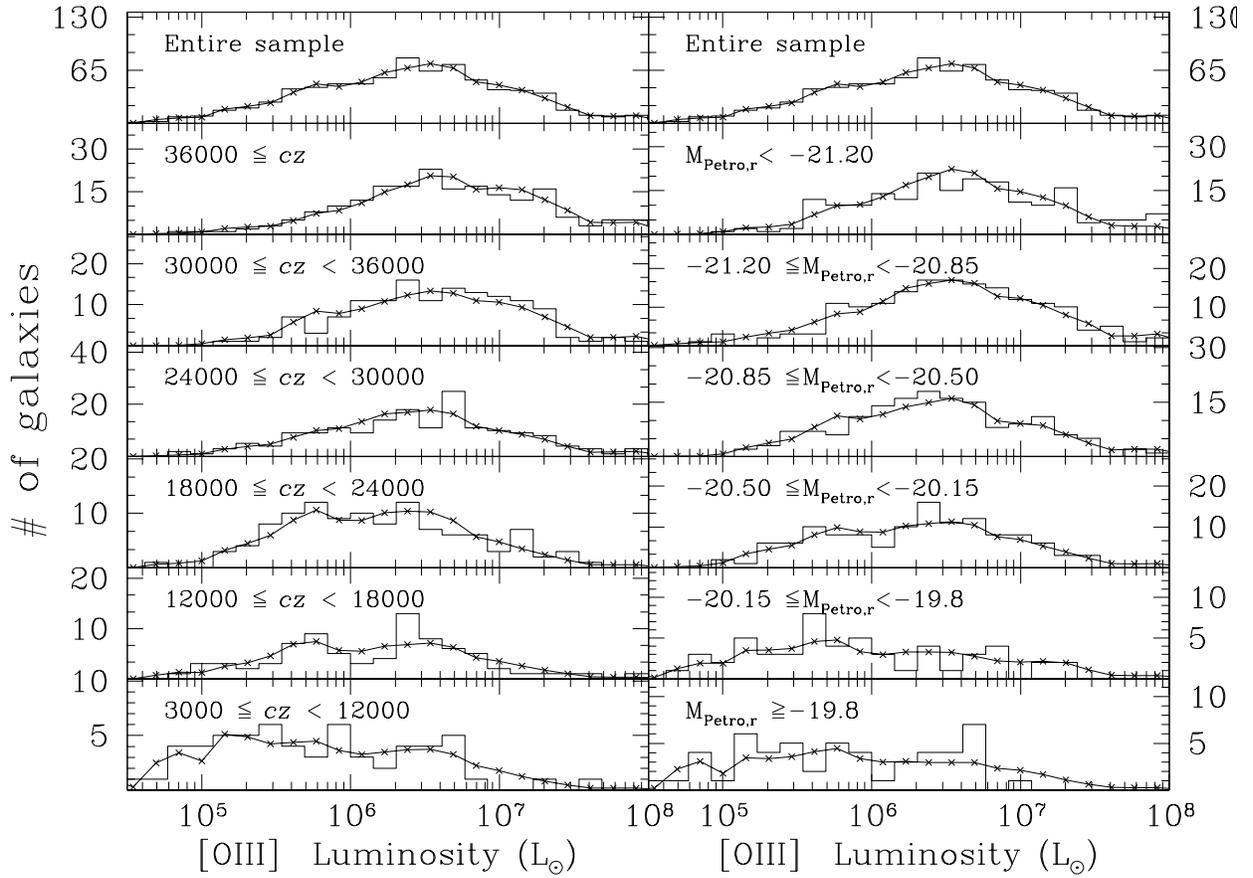}}
\caption{ The same test as in Figure~\ref{fig:lfckhann} for the Seyfert 1 [OIII] luminosity function. The match up between the observed luminosity distribution and the expected luminosity distribution in every subgroup demonstrates that the independence of nuclear luminosity and host galaxy luminosity works for broad-line AGNs as well.}
\label{fig:lfcko3b}
\end{figure}

In Figure~\ref{fig:lfcko3b}, we show the tests done for the Seyfert 1
sample via the [OIII] luminosity function. It is still a perfect match
between the observed and expected luminosity distribution for every
redshift and host galaxy magnitude subgroup. This demonstrates that
the host and AGN luminosities are not strongly correlated for Seyfert
1s either.

We apply similar tests for other luminosity functions, and always
found good match between the expected and observed luminosity
distribution.

As a further test of the correlation between host galaxy and nuclear
luminosity, we apply a correlation test to our Seyfert
samples. Correlation tests for truncated data sets have been developed
by Efron \& Petrosian (1992) and Maloney \& Petrosian (1999). Our
sample is a one-side truncated sample in flux. Fan \etal (2001)
discussed the correlation test for a one-side truncated sample with
non-sharp selection probability, which is appropriate in our study.

For each of our AGN samples, we have a one-side-truncated data set
$\left\{ L_{n,i}, L_{h,i}\right\}$, where $L_{n,i}$ and $L_{h,i}$ are
the emission line luminosity and host galaxy luminosity of the $i$th
AGN in the sample. For each AGN in the sample, we can define a
comparable data set as:
\begin{equation}
J_i=\left\{j: L_{n,j}>L_{n,i}\right\}.
\end{equation}
We denote the number of points in the set as $N_i$. Each AGN in the
data set has a detection probability $p(L_{n,i},L_{h,i})$, thus the
total number of AGN that have nuclear luminosity larger than
$L_{n,i}$, taking into account those that are missed by selection
effects, should be:
\begin{equation}
T_i=\sum_{j=1}^{N_i}\frac{p_j(L_{n,i},L_{h,j})}{p_j(L_{n,j},L_{h,j})}.
\end{equation}
Assuming that $L_{n}$ and $L_{h}$ are independent, we sort $T_i$ by
their host galaxy luminosities. The rank $R_i$ of $L_{h,i}$ among the
$T_i$ objects, which is defined as:
\begin{equation}
R_i=\sum_{j=1}^{N_i}\frac{p_j(L_{n,i},L_{h,j})}{p_j(L_{n,j},L_{h,j})}, {\rm if} L_{h,j}<L_{h,i}
\end{equation}
should be distributed uniformly between 0 and $T_i$. The expectation
value of the distribution $R_i$ is then $E_i=T_i/2$ and its variance
$V_i=T_i^2/12$. We can construct Kendall's $\tau$-statistic:
\begin{equation}
\tau=\frac{\sum(R_i-E_i)}{\sqrt{\sum V_i}}
\end{equation}
If $|\tau|<1$, the nuclear luminosity and host galaxy luminosity are
not correlated at the $1\sigma$ level and can be treated as
independent.

We assume a relationship between the host galaxy luminosity and
nuclear luminosity with a power-law index $\alpha$:
\begin{equation}
L_{h}=L_{n}^\alpha
\label{eq:tau}
\end{equation}
and run the $\left\{L_{h}, L_{n}^\alpha\right\}$ data set through the
correlation test to find the value for $\alpha$ (when $\tau=0$) and
its error (the differences of $\alpha$ when $\tau=0$ and $|\tau|=1$
). Table~\ref{tab:tau} lists the results for Seyfert 1s, Seyfert 2s
and the full Seyfert sample when $L_{nuclear}$ is represented by
[OIII] and $\Ha$ luminosities respectively. $\alpha \ll 1$ in all
cases. We conclude that the nuclear luminosity is essentially
independent of the host galaxy luminosity for all Seyferts in our
sample.
 
The above statement is obtained from our AGN sample, which
unfortunately covers only a small range of host galaxy
luminosity. Outside of this luminosity range, the independence of the
host and nuclear luminosity might not hold. For example, Kauffmann
\etal (2003) suggested that AGN of all luminosities reside almost
exclusively in massive galaxies. Within this regime, there is only a
very weak dependency of nuclear luminosity and galaxy mass, which is
in agreement with our result. But the probability of a galaxy to host
an AGN drops dramatically at galaxy masses lower than a few $\times
10^{10} M_\odot$.

\section{Comparing LFs with Previous Work}

Most previous AGN luminosity function results obtained in the
literature have been functions of B band magnitude. To make
comparisons of our luminosity function results with previous
investigators, we need to relate the emission line luminosity with the
AGN nuclear continuum in the B band. In $\S2$, we have found that
quasars with little host galaxy contamination show a tight relation
between their emission line luminosities and $r$ band PSF
magnitudes. We assume that all AGN follow the equivalent relationship
found with the $i$-band absolute magnitude:
\begin{equation}
\log (L(\Ha)/L_\odot)=-(0.470 \pm 0.011)\cdot M_i - (1.38 \pm 0.41)
\label{eq:qsohaipsf}
\end{equation}
Again, as in equation~(\ref{eq:qsohaall}), the error in the
y-intercept is chosen to include the $1\sigma$ scatter of the
relationship. With this equation, $\Ha$ luminosity function can be
converted to the $i$ band luminosity function. Based on the general
quasar color model (Schneider \etal 2002, 2003; Hopkins \etal 2004),
we found $g-i\sim 0.55$ at around $z\sim 0.1$. Also noticing that for
normal quasars, $B\sim g$, we can use the simple relationship:
\begin{equation}
B\sim i+0.55
\label{eq:Bandi}
\end{equation}
to convert the $i$ band luminosity function further to B band. 

\begin{figure}[t]
\centerline{\includegraphics[angle=-90, width=\hsize]{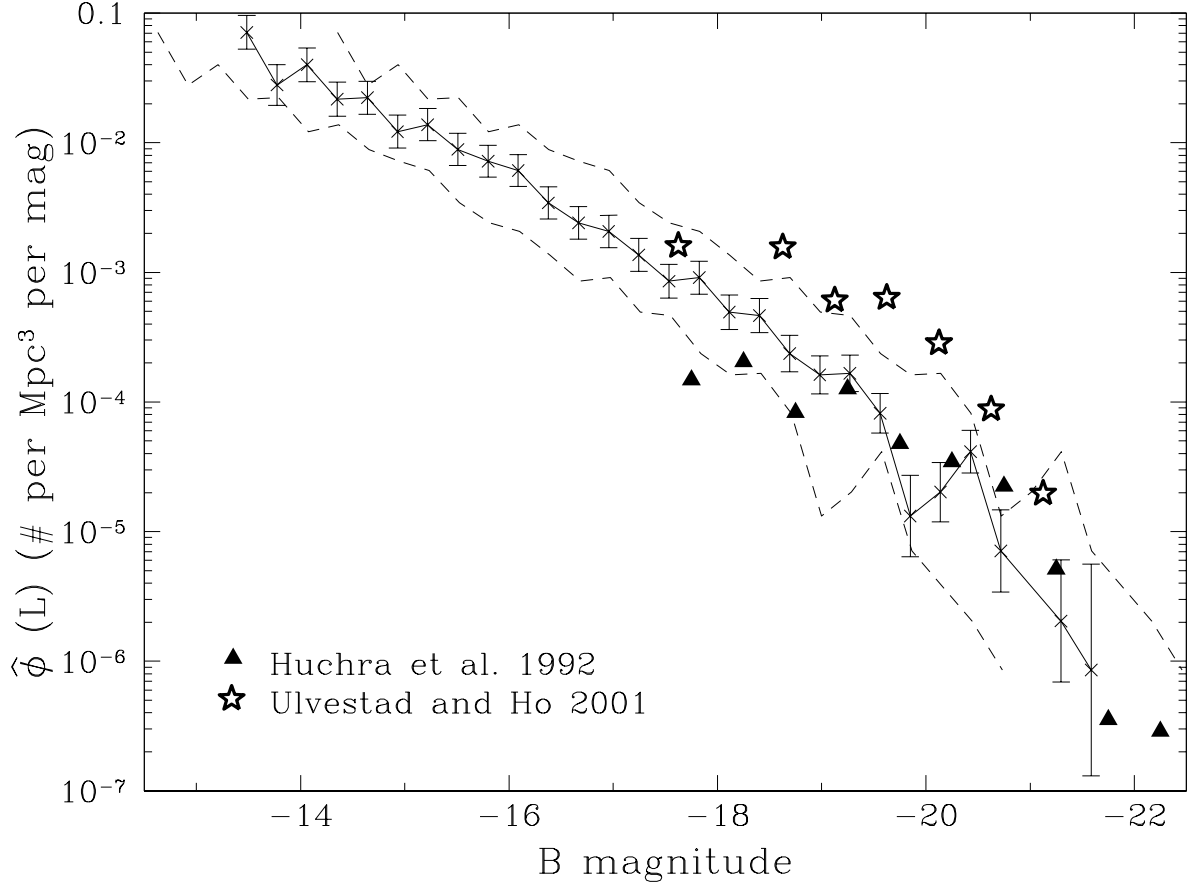}}
\caption{Our converted B band magnitude luminosity functions for Seyferts (black) with errors (dashed lines) compared to the AGN luminosity function obtained by Huchra \etal (1992) from the CfA redshift survey (solid triangles) and by Ulvestad \& Ho (2001) from the Palomar Seyfert Galaxies (stars). Our luminosity function basically agrees with the other two luminosity functions. But we should bear in mind that Huchra \etal (1992) and Ulvestad \& Ho (2001) used B band magnitudes for the whole galaxies in their luminosity function calculation, rather than the nuclear B band magnitudes we have used. }
\label{fig:lfBmaghuchra}
\end{figure}

Figure~\ref{fig:lfBmaghuchra} shows the converted B band magnitude
luminosity function for all Seyferts. The uncertainties in
equation~(\ref{eq:qsohaipsf}) are also shown (dashed line). The AGN
luminosity function results of Huchra \& Burg (1992) are plotted as
triangles, while those of Ulvestad \& Ho (2001) are plotted as
stars. Both results have been converted to $H_0=100 {\rm km\;s^{-1} \
Mpc^{-1}}$. Their choice of $(\Omega_m, \Omega_\Lambda)$ are also
different from ours, but this won't make much difference at our
redshift. In the region of overlap, these results are in good
agreement with ours, although ours reach to much fainter
luminosities. We should bear in mind that Huchra \& Burg and Ulvestad
\& Ho used host galaxy luminosities without isolating the nuclear
luminosity as we did. Therefore their luminosity function is much
flatter than ours since low-luminosity AGN can easily be buried in
host galaxy luminosities and be undetected. But since their sample
basically covers the high-luminosity end of the AGN population, their
luminosity function is not severely contaminated.

\begin{figure}[t]
\centerline{\includegraphics[angle=-90, width=\hsize]{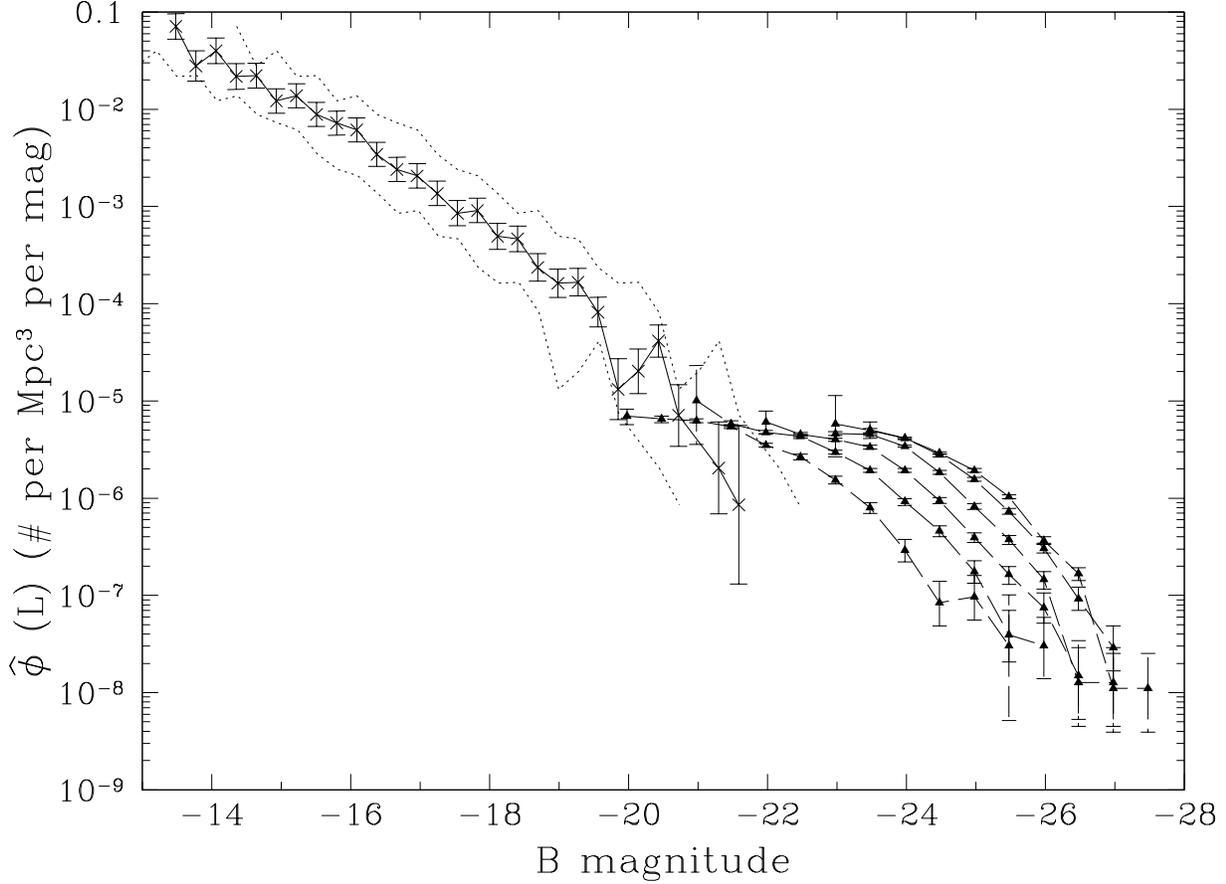}}
\caption{Our converted B band magnitude luminosity function, compared with the 2dF quasar luminosity function, converted to $H_0=100 {\rm km\;s^{-1} \, Mpc^{-1}}$. From left to right, the different 2dF quasar luminosity functions are for the redshift range: $0.40<z<0.68$, $0.68<z<0.97$, $0.97<z<1.25$, $1.25<z<1.53$, $1.53<z<1.81$ and $1.81<z<2.10$ respectively. At luminosities fainter than ${\rm M_{B}}=-21.5$, the 2dF quasar luminosity function might suffer significant incompleteness due to the effect of the host galaxy, therefore the comparison should be made only for luminosities brighter than ${\rm M_{B}}=-21.5$.
}
\label{fig:lf2dF}
\end{figure}

To compare our luminosity function results with high redshift quasar
luminosity functions, we take the 2dF QSO luminosity function obtained
by Croom \etal (2004) and converting it to our assumed cosmology. In
Figure~\ref{fig:lf2dF}, we plot our B band AGN luminosity function
(solid line with dotted lines as error range) and converted 2dF QSO
luminosity function (dashed line with triangles). Our low-redshift AGN
luminosity function agrees very well with the quasar luminosity
function for the lowest redshift range. However, our data show that
the faint end of the 2dF QSO luminosity function at low redshifts does
not flatten out. This reveals that the color selection of quasars is
subject to incompleteness at the low-luminosity range where the host
galaxy dominates (Boyle \etal 2000). Unfortunately, the large
uncertainty on our luminosity function, due to the scatter of the
conversion from the $\Ha$ luminosity to the B band magnitude, makes it
difficult to draw conclusions about AGN evolution from $z\sim 0.1$
(our result) to $z\sim 0.4$ (the lowest redshift of 2dF QSO LF). The
ongoing effort of getting the quasar luminosity function from the SDSS
will test this directly.

\section{Reddening Correction}
The luminosities measured in this paper are not corrected for any
reddening of the AGN by material in the vicinity of the AGN or the
host galaxy. However, reddening is an important factor that can
dramatically change the observed luminosity of an AGN. The standard
method of reddening correction is to measure the Balmer decrements and
assume an intrinsic $\Ha/\Hb$ value (\eg 2.86) and a reddening
function (\eg Seaton 1979; Charlot \& Fall 2000). However, AGN with
large Balmer decrements have been observed before (Anderson 1970;
Baldwin 1975; Weedman 1977; Rix \etal 1990; Goodrich 1990; Barcons
\etal 2003), many of these can not be simply explained by dust
obscuration. Instead, AGN might have a higher intrinsic $\Ha/\Hb$
ratio, probably because of higher densities and radiative transfer
effects. In addition, the relative distribution of dust and emission
line gas can be quite complicated, while most models assume that the
dust lies in a screen in front of the line-emitting regions. Due to
these difficulties and uncertainties, we believe it is not appropriate
to do a standard reddening correction using observed Balmer
decrements, even though reddening is clearly present. The whole issue
of AGN obscuration needs much further investigation and
multiwavelength observations from infrared and X-ray, and is beyond
the scope of this paper. We will leave the luminosity function derived
in this paper reddening uncorrected until reliable and accurate AGN
reddening correction methods are developed.

\section{Summary and Discussion}
Using the AGN sample obtained from Paper I, we have evaluated the AGN
[OIII] and $\Ha$ emission line luminosity functions. Because of the
complexity of our AGN selection criteria, we use a Monte-Carlo
simulation for each AGN to obtain its selection probability
function. The overall $\Ha$ emission line luminosity function for
Seyferts covers a broad range of luminosity and approximately follows
a two power-law form. After converting to B band magnitude luminosity
functions, we found that our results are comparable to the luminosity
function obtained by Huchra \etal (1992) from the CfA redshift survey
and Ulvestad \& Ho (2001) from the Palomar spectroscopic survey of
nearby galaxies. It also agrees well with the low-redshift bin of the
2dF QSO luminosity function by Croom \etal (2004). By integrating the
overall $\Ha$ luminosity function from $10^5$ to $10^9$ $L_\odot$, we
obtain the AGN number density of 0.018 Mpc$^{-3}$, approximately 20\%
of all galaxies. We also estimated the AGN $r$ band luminosity density
as $1.11\times 10^6 L_\odot {\rm Mpc}^{-3}$, which is about $6\times
10^{-3}$ of the luminosity density of galaxies.

The comparison of the [OIII] luminosity functions for Seyfert 1s and
Seyfert 2s reveals that the relative number ratio of the two types of
AGN is a function of luminosity: at low luminosity, the ratio of
Seyfert 1s to Seyfert 2s is about one, while at high luminosities,
Seyfert 1s outnumber Seyfert 2s. The conclusion is not sensitive to
the uncertainties in the selection criteria. The paucity of
narrow-line objects in high luminosity AGN is unexpected in the
simplest unification model, and various models have been proposed to
accommodate it. For example, Lawrence (1991) proposed that the opening
angle of the dust torus is larger for more luminous AGN, presumably
because increased luminosity means more dust sublimation, and thus the
broad-line region can be seen over a larger opening angle. K\"{o}nigl
\& Kartjel (1994) proposed a disk-wind AGN model, in which the
radiation pressure force flattens the dust distribution in objects
with comparatively high bolometric luminosities, thus the opening
angle of the dust torus becomes larger for more luminous AGN. Our
result for the first time gives the opportunity to determine the
detailed opening angle as a function of luminosity and will provide
detailed constraints to various models. This will be fully
investigated in future papers. To fully understand the physics behind
the lack of Seyfert 2s at high luminosity, multiwavelength
observations of AGN, especially the infrared will be necessary. By
directly measuring dust emission in Seyfert 1s and Seyfert 2s, various
dust properties can be determined, and will help to understand if dust
is playing a major role in the difference between Seyfert 1s and
Seyfert 2s.

Our AGN luminosity function is obtained by assuming that the AGN
nuclear luminosities are independent of the host galaxy
luminosities. Detailed verification of this assumption has been
carried out and we show strong evidence that the two variables are not
strongly correlated. Assuming that AGN are hosted by normal galaxies,
their host galaxy bulge luminosities are well correlated with their
velocity dispersions, which in turn are strongly correlated with the
central black hole masses (Gebhardt \etal 2000; Ferrarese \& Merritt
2000; Ferrarese \etal 2001; Tremaine \etal 2002). The independence of
the nuclear luminosity and the host galaxy luminosity implies that the
nuclear luminosity does not depend on the central black hole mass
(Heckman \etal 2004). Therefore, for the AGN in our sample, the
principal variable that determines the nuclear luminosity is the
Eddington ratio, defined as $\lambda= L_{bol}/L_{Eddington}$, instead
of the black hole mass.

The independence of AGN nuclear luminosities and their central black
hole masses is supported by several observations. For example, Woo \&
Urry (2002) compiled a sample of 377 AGN with known black hole masses
and calculated bolometric luminosities for 234 AGN among them. When
relating these bolometric luminosities with their black hole masses,
they found no significant correlation between the two
variables. O'Dowd \& Urry (2002) also reported a very weak trend
between the nuclear and host luminosities for a sample of radio loud
AGN.

However, the same statement does not seem to work for high-redshift
quasars. A black hole mass inferred from the observed luminosity is a
lower limit. Thus given their high luminosities, one would infer
implausibly high black hole mass ($>10^{14}M_\odot$) if the most
luminous quasars are accreting at substantially sub-Eddington
rates. (see Vestergaard \etal 2002 for black hole mass measurements
for high-redshift quasars). Thus, it will be interesting to explore
the transition from low to high Eddington rate accretion (McLure \& Dunlop
2004). Hopefully, the ongoing evaluation of the quasar luminosity
function obtained from the SDSS will yield insights on this issue.

Our result will also bring insights to X-ray background and UV
background problems. Detailed investigation is beyond the scope of the
current paper. The multiwavelength observation of the sample will be
explored in future papers.

{\bf Acknowledgments.} \,\,\,We would like to thank Lisa Kewley for
useful discussions and comments and Scott Croom for providing the
latest 2dF quasar luminosity function values.

Funding for the creation and distribution of the SDSS Archive has been
provided by the Alfred P. Sloan Foundation, the Participating
Institutions, the National Aeronautics and Space Administration, the
National Science Foundation, the U.S. Department of Energy, the
Japanese Monbukagakusho, and the Max Planck Society. The SDSS Web site
is http://www.sdss.org/.

The SDSS is managed by the Astrophysical Research Consortium (ARC) for
the Participating Institutions. The Participating Institutions are The
University of Chicago, Fermilab, the Institute for Advanced Study, the
Japan Participation Group, The Johns Hopkins University, the Korean
Scientist Group, Los Alamos National Laboratory, the
Max-Planck-Institute for Astronomy (MPIA), the Max-Planck-Institute
for Astrophysics (MPA), New Mexico State University, University of
Pittsburgh, Princeton University, the United States Naval Observatory,
and the University of Washington.

L.H. and M.A.S. are supported in part by NSF grant AST-0071091 and AST-0307409. L.H. acknowledges the support of the Princeton University Research Board.

\clearpage

\clearpage

\clearpage

\clearpage

\clearpage

\clearpage

\clearpage

\clearpage

\clearpage

\clearpage

\clearpage

\clearpage

\clearpage
\begin{deluxetable}{lccccccccccccccc} 
\rotate

\tablecolumns{16}
\tablewidth{0pc} 
\tabletypesize{\scriptsize}

%\tablecaption{\sc{Fitting Results}} \label{tab:fit}
\tablecaption{\sc{Fitting Results}} 
\tablehead{ 
\colhead{}    &  \multicolumn{4}{c}{Schechter function} &   \colhead{}   & 
\multicolumn{5}{c}{two-powerlaw function} & \colhead{}   & 
\multicolumn{4}{c}{one-powerlaw function}\\ 
\cline{2-5} \cline{7-11} \cline{13-16}\\ 
\colhead{AGN sample} & \colhead{$\phi^*$}   & \colhead{$L^*$}    & \colhead{$\alpha$} &  \colhead{$\chi^2$/DOF}   &
 \colhead{}    & \colhead{$\phi^*$}   & \colhead{$L^*$}    & \colhead{$\alpha$}  &   \colhead{$\beta$}   & \colhead{$\chi^2$/DOF}   &
 \colhead{}    & \colhead{$\phi^*$}   & \colhead{$L^*$}    & \colhead{$\alpha$}  &  \colhead{$\chi^2$/DOF} }

\startdata 
All Seyfert $\Ha$ &$9.17\times 10^{-6}$ &$3.16\times 10^8$ &-1.94 &17.3/25 &  &$4.11\times 10^{-5}$ &$8.51\times 10^7$ &2.78 &1.88 &16.7/24 &  &$3.04\times 10^{-5}$ &$5.97\times 10^7$ &-2.02 &26.8/25\\ 
Sey 2 $\Ha$ &$2.49\times 10^{-5}$ &$2.59\times 10^7$ &-1.90 &58.9/18 &  &$1.03\times 10^{-4}$ &$7.40\times 10^6$ &2.91 &1.77 &44.7/17 &  &$1.12\times 10^{-5}$ &$3.01\times 10^7$ &-2.05 &119.5/18\\ 
Sey 1 $\Ha$ &$3.05\times 10^{-6}$ &$5.74\times 10^8$ &-2.02 &31.1/23 &  &$3.37\times 10^{-6}$ &$4.69\times 10^8$ &5.12 &2.05 &30.0/22 &  &$2.75\times 10^{-5}$ &$5.67\times 10^7$ &-2.07 &33.0/23\\
Sey 1 $\Ha$ narrow &$9.17\times 10^{-6}$ &$1.25\times 10^8$ &-1.87 &32.2/25 &  &$2.01\times 10^{-4}$ &$8.78\times 10^6$ &2.66 &1.54 &18.6/24 &  &$3.04\times 10^{-5}$ &$2.11\times 10^7$ &-2.02 &55.2/25\\ 
Sey 2$^*$ $\Ha$ &$3.04\times 10^{-5}$ &$5.29\times 10^7$ &-2.06 &115.0/20 &  &$4.73\times 10^{-5}$ &$3.13\times 10^7$ &4.68 &2.08 &105.1/19 &  &$2.75\times 10^{-5}$ &$4.53\times 10^7$ &-2.12 &122.3/20\\
All Seyfert [OIII] &$6.19\times 10^{-9}$ &$2.13\times 10^{10}$ &-2.15 &226.5/24 &  &$3.72\times 10^{-5}$ &$1.99\times 10^7$ &2.16 &2.15 &226.3/23 &  &$1.12\times 10^{-5}$ & $3.24\times 10^7$ &-2.15 &226.3/24\\ 
Sey 2 [OIII] &$1.86\times 10^{-8}$ &$1.55\times 10^{9}$ &-2.35 &213.0/22 &  &$1.68\times 10^{-4}$ &$2.96\times 10^6$ &2.35 &2.35 &210.1/21 &  &$8.29\times 10^{-6}$ & $1.67\times 10^7$ &-2.35 &210.1/22\\
Sey 1 [OIII] &$1.02\times 10^{-6}$ &$3.73\times 10^{8}$ &-2.04 &56.0/18 &  &$2.04\times 10^{-5}$ &$3.72\times 10^7$ &2.07 &2.08 &58.7/17 &  &$1.37\times 10^{-5}$ & $2.86\times 10^7$ &-2.07 &58.7/18\\ 
Sey 2$^*$ [OIII] &$4.13\times 10^{-7}$ &$4.70\times 10^{8}$ &-2.14 &62.5/15 &  &\nodata &\nodata &\nodata &\nodata &\nodata &  &$9.17\times 10^{-6}$ & $3.03\times 10^7$ &-2.15 &62.1/15\\
Sey 2 [OII] &$1.36\times 10^{-4}$ &$5.17\times 10^{6}$ &-1.58 &104.0/17 &  &$3.63\times 10^{-4}$ &$1.97\times 10^6$ &3.23 &1.29 &35.4/16 &  &$1.37\times 10^{-5}$ & $2.07\times 10^7$ &-1.97 &294.7/17\\ 
Sey 1 [OII] &$7.34\times 10^{-5}$ &$1.18\times 10^{7}$ &-1.59 &36.7/17 &  &$6.17\times 10^{-4}$ &$1.83\times 10^6$ &2.94 &0.83 &14.5/16 &  &$6.79\times 10^{-6}$ & $3.54\times 10^7$ &-2.03 &64.4/17\\ 
 Sey 2$^*$ [OII] &$4.53\times 10^{-4}$ &$4.84\times 10^{6}$ &-1.62 &57.9/9 &  &$4.21\times 10^{-3}$ &$8.74\times 10^5$ &3.29 &0.55 & 8.1/8 &  &$1.24\times 10^{-5}$ & $3.94\times 10^7$ &-2.10 &104.1/9\\ 

\enddata
\tablecomments{$\phi^*$ is in unit of ${\rm Mpc}^{-3}$, and $L^*$ is in unit of $L_\odot$}
\label{tab:fit}
\end{deluxetable}

\clearpage
\begin{deluxetable}{lcr}
%\tabletypesize{\scriptsize}
\tablewidth{0pt}
\tablecaption{{\sc{Correlation Test Results}} \label{tab:tau}}
\tablehead{
\colhead{\sc{AGN Sample}} & \colhead{$L_{nuclear}$}   &
\colhead{$\alpha$ ($L_{host}=L_{nuclear}^\alpha$)}
}
\startdata  
Seyfert 1s  \dotfill & [OIII] & 0.0088 $\pm$ 0.0012\\
Seyfert 2s  \dotfill & [OIII] & -0.0029 $\pm$ 0.0009\\
All Seyferts  \dotfill & [OIII] & 0.0003 $\pm$ 0.0007\\
Seyfert 1s  \dotfill & $H\alpha$ & 0.0094 $\pm$ 0.0011\\
Seyfert 2s  \dotfill & $H\alpha$ & 0.0073 $\pm$ 0.0008\\
All Seyferts  \dotfill & $H\alpha$ & 0.0043 $\pm$ 0.0006\\

\enddata
\end{deluxetable}

\end{document}